\documentclass{aa}  

\usepackage{graphicx}
\usepackage{txfonts}
\usepackage{natbib}
\usepackage{pifont}
\usepackage{amsmath}
\usepackage{graphicx}
\usepackage{txfonts}
\usepackage{url}
\usepackage{subfigure}
\usepackage{longtable}
\usepackage{lscape}
\usepackage[colorlinks,linkcolor=red,anchorcolor=green,citecolor=blue]{hyperref}
\usepackage[normalem]{ulem}

\bibpunct{(}{)}{;}{a}{}{,}

\newcommand{\msol}{\mbox{M$_\odot$}}
\newcommand{\mjup}{\mbox{M$_{\rm Jup}$}}

\newcommand{\lsol}{\mbox{L$_\odot$}}

\newcommand\adeg{\mbox{$^\circ$}}%

\newcommand{\dotsec}{\rlap.{''}}
\newcommand{\dotmin}{\rlap.{'}}

\newcommand{\au}{\rm au}	    
\newcommand {\asp}{\mbox{$.\!\!^{\prime\prime}$}}

\newcommand{\kms}{\mbox{km\,s$^{-1}$}}

\begin{document}

\title{A resolved rotating disk wind from a young T\,Tauri star in the Bok globule CB\,26
      \thanks{Based on observations carried out with the IRAM Plateau de Bure Interferometer (PdBI, now NOEMA), the Owens Valley millimeter-wave array (OVRO), and the Submillimeter Array (SMA).}}

\author{R. Launhardt \inst{1},
Ya. N. Pavlyuchenkov \inst{2}, 
V. V. Akimkin \inst{2},
A. Dutrey \inst{3,4},
F. Gueth \inst{5},
S. Guilloteau \inst{3,4},
Th. Henning \inst{1},
V. Pi\'etu \inst{5}, 
K. Schreyer \inst{6},
D. Semenov \inst{1},
B. Stecklum \inst{7},
T. L. Bourke \inst{8}
}
%
%
\institute{Max-Planck-Institut f\"ur Astronomie, K\"onigstuhl 17, D-69117
  Heidelberg, Germany 
\and
  Institute of Astronomy, Russian Academy of Sciences, Pyatnitskaya 48, Moscow, 119117, Russia
\and
  Laboratoire d’Astrophysique de Bordeaux, All\'ee Geoffroy Saint Hilaire, 33600, Pessac, France
\and
  CNRS/INSU - UMR5804; BP 89, F-33270 Floirac, France
\and
  IRAM, 300 rue de la piscine, F-38406 Saint Martin d'H\`eres, France
\and 
  Astrophysikalisches Institut und Universit\"ats-Sternwarte, Schillerg\"asschen 2-3, D-07745 Jena, Germany
\and
  Th\"uringer Landessternwarte Tautenburg, Sternwarte 5, D-07778 Tautenburg, Germany
\and
  SKA Observatory, Jodrell Bank, Lower Withington, Macclesfield SK11 9FT, United Kingdom.
}
\date{Received 17 July 2023 / Accepted 31 August 2023 }



\abstract
{The disk-outflow connection plays a key role in extracting excess angular momentum from a forming protostar. Although indications of jet rotation have been reported for a few objects, observational constraints of outflow rotation are still very scarce. We have previously reported the discovery of a small collimated molecular outflow from the  edge-on T\,Tauri\,star\,--\,disk system in the Bok globule CB\,26 that shows a peculiar velocity pattern, reminiscent of an outflow that corotates with the Keplerian disk. However, we could not ultimately exclude possible alternative explanations for the origin of the observed velocity field.}
{We report new, high angular resolution millimeter-interferometric observations of CB\,26 with the aim of revealing the morphology and kinematics of the outflow at the disk\,--\,outflow interface to unambiguously discriminate between the possible alternative explanations for the observed peculiar velocity pattern.}
{The IRAM PdBI array and the 30\,m telescope were used to observe HCO$^{+}$(1--0) and H$^{13}$CO$^{+}$(1--0) at 3.3\,mm and \mbox{$^{12}$CO(2--1)} at 1.3\,mm in three configurations plus zerospacing, resulting in spectral line maps with angular resolutions of 3\farcs5 and 0\farcs5, respectively. The SMA was used to observe the HCO$^{+}$(3--2) line at 1.1\,mm with an angular resolution of 1\farcs35. Additional earlier observations of $^{13}$CO(1--0) at 2.7\,mm with an angular resolution of 1\farcs0, obtained with OVRO, are also used for the analysis. Using a physical model of the disk, which was derived from the dust continuum emission, we employed chemo-dynamical modeling combined with line radiative transfer calculations to constrain kinematic parameters of the system and to construct a model of the CO emission from the disk that allowed us to separate the emission of the disk from that of the outflow.}
{Our observations confirm the disk-wind nature of the rotating molecular outflow from \mbox{CB\,26\,-\,YSO\,1}. The new high-resolution data reveal an X-shaped morphology of the CO emission close to the disk, and vertical streaks extending from the disk surface with a small half-opening angle of $\approx$7\degr, which can be traced out to vertical heights of $\approx$500\,au. We interpret this emission as the combination of the disk atmosphere and a well-collimated disk wind, of which we mainly see the outer walls of the outflow cone. The decomposition of this emission into a contribution from the disk atmosphere and the disk wind allowed us to trace the disk wind down to vertical heights of $\approx$40\,au, where it is launched from the surface of the flared disk at radii of $R_{\rm L}\approx$\,20\,--\,45\,au. 
The disk wind is rotating with the same orientation and speed as the Keplerian disk and the velocity structure of the cone walls along the flow is consistent with angular momentum conservation.
The observed CO outflow has a total gas mass of $\approx10^{-3}$\,\msol, a dynamical age of \mbox{$\tau_{\rm dyn}\approx740$\,yr}, and a total momentum flux of $\dot{P}_{\rm CO}\approx 1.0\times10^{-5}$\,\msol\,\kms\,yr$^{-1}$, which is nearly three orders of magnitude larger than the maximum thrust that can be provided by the luminosity of the central star.}
{We conclude that photoevaporation cannot be the main driving mechanism for this outflow, but it must be predominantly a magnetohydrodynamic (MHD) disk wind. It is thus far the best-resolved rotating disk wind observed to be launched from a circumstellar disk in Keplerian rotation around a low-mass young stellar object (YSO), albeit also the one with the largest launch radius. It confirms the observed trend that disk winds from Class\,I YSOs with transitional disks have much larger launch radii than jets ejected from Class\,0 protostars.}

%
%
%
\keywords{circumstellar matter --- ISM: jets and outflows --- 
    Stars: pre-main sequence --- planetary systems: protoplanetary disks ---
    ISM: molecules}

\titlerunning{The CB\,26 rotating disk wind}
\authorrunning{R. Launhardt et al.}

\maketitle

\section{Introduction} \label{sec:intro}

Similar to other accretion processes in the universe, such as on black holes at the centers of radio galaxies, the formation of stars and planetary systems is always accompanied by accretion disks and outflows. The theory of magnetohydrodynamic (MHD) winds launched from magnetized accretion disks was first outlined by \citet{blandford1982}, and applied to circumstellar disks around young stars by \citet{pudritz1983}. These winds are predicted to be launched from most of the disk surface, efficiently carry away angular momentum from the disk, and should therefore rotate. They enable accretion from the disk onto the forming star at a rate of $\dot{M}_{\rm of}/\dot{M}_{\ast}\approx0.1\ldots0.2$, where $\dot{M}_{\rm of}$\ is the mass outflow rate, and $\dot{M}_{\ast}$\ is the mass accretion rate onto the star \citep[e.g.,][]{pudritz1986,watson2016,lee2020}. An alternative mechanism was proposed by \citet{shu1994}, in which high-velocity outflows are launched from the corotation radius of the star and the disk, where the star’s magnetosphere interacts directly with the inner edge of the disk, the so-called X-point \citep[see also][]{Shu2000}. Similar to the disk wind models, this mechanism also predicts that jets and outflows carry away angular momentum and should thus rotate. 

While rotation signatures in protoplanetary disks have been observed since the advent of radio (millimeter) interferometers \citep{sargent1987}, and since then they have been observed and spatially resolved more routinely \citep[e.g.,][]{Simon2000,launhardt01,Qi2003,panic2009,oeberg2010,huang2021}, rotation signatures in outflows seem to be notoriously more difficult to detect. Transverse velocity gradients have been observed across a number of protostellar high-velocity jets using high-resolution NASA Hubble Space Telescope (HST) spectra \citep[e.g.,][]{bacciotti02,bacciotti03,coffey04,coffey2007,woitas_etal2005} and more recently by, for example, \citet{lee2017}. Rotation signatures have now also been found in a number of slow molecular winds and outflows \citep[e.g.,][]{launhardt09,lee2009,klaassen2013,chen2016,bjerkeli2016,hirota2017}. Recent reviews on the theory and observations of protostellar jets, outflows, as well as disk winds can be found in \citet{pudritz2019}, \citet{lee2020}, and \citet{pascucci2023}. 

In this paper, we present new data and analyze the physical properties of the disk and outflow related to the young T\,Tauri star (TTS) embedded in CB\,26. CB\,26 (L\,1439) is a small cometary-shaped, double-core Bok globule located $\approx$\,10$^\circ$ north of the Taurus-Auriga dark cloud, at a distance of $140\pm20$\,pc\footnote{\citet{das2015} find a photometrically estimated distance of 293$\pm54$\,pc to the nearby globule CB\,24, which is most likely physically related to CB\,26. We do not use their distance because their findings are not supported by the {\it Gaia} data.} \citep{launhardt10,loinard2011,launhardt13}. While the eastern sub-core is starless, a second dense core with signatures of star formation is located at the south west rim of the globule \citep{stutz2009,launhardt13}. 
OVRO observations of the millimeter dust continuum emission and of the $^{13}$CO\,(1-0) line have revealed a nearly edge-on circumstellar disk of radius $\approx$200\,au with Keplerian rotation, surrounding a very young ($\le$\,1\,Myr) low-mass ($\approx$\,0.5\,\msol) TTS \citep{launhardt01}. It is associated with a small bipolar near-infrared (NIR) nebula that is bisected by a dark extinction lane at the position and orientation of the edge-on disk \citep{stecklum04}. This disk and its associated near-infrared (NIR) nebula are referred to as \mbox{CB\,26\,-\,YSO\,1}\footnote{This naming was chosen to be consistent with other globule-related papers, although no second YSO has been identified in CB\,26 (yet).} in the following. The source is surrounded by an optically thin asymmetric envelope with a well-ordered magnetic field directed along P.A.\,$\approx$\,25\degr\ \citep{henning01}. Furthermore, a Herbig-Haro (HH) object  was identified by H$\alpha$ and S[II] narrow-band imaging, 6\dotmin15 north west of CB\,26 at P.A.\,$\approx145$\degr\ \citep[HH\,494,][]{stecklum04}. Figure\,\ref{fig-overview} shows an overview of CB\,26 and its immediate surroundings.

Using the PdBI, we have detected a small ($2\times 2000$\,AU) collimated bipolar molecular outflow from this source, which shows peculiar kinematic signatures. We used an empirical steady-state outflow model combined with 2-D line radiative transfer calculations and $\chi^2$-minimization to derive a best-fit model and constrain parameters of the outflow \citep{launhardt09}. We could show that the data are best reproduced by an outflow that is rotating about its polar axis. This hypothesis was supported by the fact that disk and outflow are corotating, fit together energetically, and by the presence of HH\,494, which is located $\approx$0.25\,pc away from CB\,26 but is exactly aligned with the outflow axis \citep{stecklum04}. Although we could not ultimately exclude alternative scenarios such as jet precession or two misaligned jets from a hypothetical embedded binary system as possible origins of the observed peculiar velocity field, CB\,26 was at this time the most promising source in which to study the dispersion of disk angular momentum by a rotating molecular outflow. More recently, \citet{lopez2022} have shown that the rotating outflow can indeed be well-explained by a disk wind.

To verify the hypotheses of outflow rotation and disk wind origin in CB\,26 against the possible alternative scenarios mentioned above and in \citet{launhardt09}, we obtained higher-resolution $^{12}$CO(2--1) data, which we present and analyze in this paper. Since the CO emission from the disk atmosphere and the outflow are not separated spatially at the origin of the outflow, we establish a physical disk model based on radiative transfer models of the dust continuum emission, use the results to model the CO emission from the disk, and subtract it from the data.

This paper is structured as follows:
In Section\,\ref{sec:obs} we describe the observations and data reduction.
The direct results on the disk and outflow are presented in Section\,\ref{sec:res}.
In Section\,\ref{sec:mod}, we present the disk model and analyze the disk-subtracted CO emission from the outflow. 
These results, their uncertainties, and the physical implications are discussed in Section\,\ref{sec:dis}.
Finally, Section\,\ref{sec:sum} summarizes the paper.


\section{Observations and data reduction} \label{sec:obs}

CB\,26\,-\,YSO\,1 has been observed at many wavelengths, ranging from 0.9\,$\mu$m to 6.4\,cm. The thermal dust continuum observations and data reduction have been described in detail and the photometry over the entire spectral range is listed in \citet{zhang2021}. Here we only describe the molecular line observations and the optical and near-infrared (NIR) observations used in this paper.

\begin{table*}[ht]
\caption{List of millimeter molecular line observations of CB\,26\tablefootmark{a} and summary of observing parameters.}
\label{tab-obs} 
{\footnotesize  
\begin{tabular}{lccccccccc}
\hline \hline
Line & Frequ. & Tel. & Year & {\it uv} radius & Bandw.         & $\Delta v_{\rm chan}$ & Primary  & Synthesized    & 1\,$\sigma$\,rms \\
     &        &      &      & range           &                &                       & HPBW     & HPBW (PA)      &                  \\
     & [GHz]  &      &      & [m]             & [\kms] & [\kms]        & [arcsec] & [arcsec (deg)] & [mJy/beam]       \\
\hline 
$^{12}$CO\,(2--1)       & 230.538  & PdBI\tablefootmark{a}      & 2005,2009 & 18\tablefootmark{b}\,--\,753 & 14  & 0.20 & 21.9 & 0.53$\times$0.47 (6.9)\tablefootmark{c}   & 8 \\
$^{13}$CO\,(1--0)       & 110.201  & OVRO      & 2001      & 12\,--\,478 & 4.8 & 0.15 & 45.7 & 1.20$\times$0.88 (91.7)  & 35 \\
HCO$^+$\,(1--0)         & ~~89.189 & PdBI\tablefootmark{a}      & 2005      & 17\tablefootmark{b}\,--\,175 & 12.8 & 0.20 & 56.5 & 3.74$\times$3.38 (100.3)  & 7  \\
H$^{13}$CO$^+$\,(1--0)  & ~~86.754 & PdBI\tablefootmark{a}      & 2008,2009 & 20\tablefootmark{b}\,--\,175 & 7    & 0.50 & 58.1 & 4.09$\times$2.26 (114.6) & 5  \\
HCO$^+$\,(3--2)         & 267.558  & SMA       & 2006      & 11\,--\,192 & 29  & 0.23 & 18.8 & 1.37$\times$1.34 (39.0)   & 150 \\ 
\hline
\end{tabular} 
\tablefoot{
\tablefoottext{a}{The phase center was at $\alpha_{2000} = 04^h59^m50.74^s$, $\delta_{2000} = 
           52^{\circ}04^{\prime}43.80^{\prime\prime}$.}
\tablefoottext{b}{Plus zero-spacing observations with the IRAM 30\,m telescope (see Sect.\,\ref{sec:obs:pdbi}).}
\tablefoottext{c}{Natural $uv$-weighting. We also produced maps with robust $uv$-weighting (HPBW 0.48$\times$0.42\arcsec) 
                  and a tapered (smoothed) map (HPBW 1.08$\times$0.92\arcsec, see Fig.\,\ref{fig_covelfield}).}
}}
\end{table*}
\begin{table*}[ht]
\caption{Purpose and use of the millimeter molecular lines.}
\label{tab-lines} 
{\footnotesize  
\begin{tabular}{lclclll}
\hline \hline
Line & Frequ. & $n_{\rm crit}$\,(10\,K) & Ref. & Advantages & Disadvantages & Used for\tablefootmark{a} \\
     & [GHz]  & [cm$^{-3}$]             &      &            &               &                           \\
\hline 
$^{12}$CO\,(2--1)       & 230.538  & $1\times10^3$   & 1 & Abundant, low $n_{\rm crit}$ & Often optically thick,  & Mass estimate of outflow,   \\
&   &   &   & best (only?) tracer of outflow  & Self-absorption by envelope  & velocity field of disk and outflow      \\
$^{13}$CO\,(1--0)       & 110.201  & $8\times10^2$   & 1 & Low $n_{\rm crit}$ & Does not trace the outflow    & Velocity field of disk     \\
                        &          &                 &   &  mostly optically thin  & Self-absorption by envelope  &  \\
HCO$^+$\,(1--0)         & ~~89.189 & $4.5\times10^4$ & 2 &  High $n_{\rm crit}$, strong line & Traces only inner part of disk & Velocity field of inner disk     \\
                        &          &                 &   &     & Self-absorption by envelope &    \\
H$^{13}$CO$^+$\,(1--0)  & ~~86.754 & $4.1\times10^4$ & 2 & High $n_{\rm crit}$, optically thin  & Barely detected    & Not used     \\
                        &          &                 &   &             &              &      \\
HCO$^+$\,(3--2)         & 267.558  & $1.4\times10^6$ & 2 &  High $n_{\rm crit}$, not excited in env. & Traces only inner part of disk              & Velocity field of inner disk     \\
                        &          &                 &   &  No self-abs. by envelope &              &   \\
\hline
\end{tabular} 
\tablefoot{
\tablefoottext{a}{We only list the use in this paper. For other targets and conditions, these lines may also be used for other purposes.}}
\tablebib{
(1)~\mbox{\citet{beslic2021}};
(2)~\mbox{\citet{Shirley2015}}
}
}
\end{table*}


\subsection{OVRO millimeter observations} \label{sec:obs:ovro}

CB\,26 was observed with the Owens Valley Radio Observatory (OVRO)
between January 2000 and December 2001. Four configurations of the six 10.4\,m antennas provided baselines in the range 6\,--\,180\,k$\lambda$\ at 2.7\,mm (110\,GHz). Average SSB system temperatures of the SIS receivers were 300\,--\,400\,K. The digital correlator was centered on the \mbox{$^{13}$CO(1--0)} line at 110.2\,GHz, adopting the systemic velocity\footnote{All radial velocity values in this paper refer to the Local Standard of Rest (LSR).} of CB\,26, \mbox{$v_{\rm LSR} = 5.95$\,\kms}. Spectral resolution and bandwidth were 0.15\,\kms\ and 4.8\,\kms, respectively. The OVRO observations and data are described in more detail in \citet{launhardt01}. The data presented here include additional \mbox{$^{13}$CO(1--0)} observations conducted in 2001. 
The OVRO raw data were calibrated and edited using the MMA software package \citep{scoville93}. After mapping and checking the results for all individual data sets, the calibrated continuum {\it uv} tables were combined and imaged together with the respective PdBI {\it uv} tables as described below. The $^{13}$CO line data were also imaged with GILDAS in the same way as the PdBI line data.


\subsection{IRAM PdBI and 30\,m single-dish millimeter observations} \label{sec:obs:pdbi}

Observations of \mbox{HCO$^{+}$\,(1--0)} and \mbox{$^{12}$CO\,(2--1)} in CB\,26 were carried out with the IRAM PdBI in November 2005 (D configuration with 5 antennas) and December 2005 (C configuration with 6 antennas; project PD0D). Two receivers were used simultaneously and tuned single side-band (SSB) to the \mbox{HCO$^{+}$\,(1--0)} line at 89.188526\,GHz, and the $^{12}$CO\,(2-1) line at 230.537984\,GHz, respectively, adopting the systemic velocity of CB\,26. 
Further observations of \mbox{H$^{13}$CO$^{+}$\,(1--0)} where carried out in November 2008 (C configuration) and in March 2009 (D configuration; project SC1C), with 6 antennas and using the new 3\,mm receivers. Additional higher-resolution observations of \mbox{$^{12}$CO\,(2--1)} at 230.5\,GHz were carried out in January 2009 (B configuration) and in February 2009 (A configuration with 6 antennas; project S078).
Several nearby phase calibrators were observed during all tracks to determine the time-dependent complex antenna gains. The correlator bandpasses were calibrated on 3C454.3 and 3C273, and the absolute flux density scale was derived from observations of MWC\,349.  The flux calibration uncertainty is estimated to be $\leq$20\% at both wavelengths. The phase center was at \mbox{$\alpha_{2000} = 04^h59^m50.74^s$}, \mbox{$\delta_{2000} = 52^{\circ}04^{\prime}43.80^{\prime\prime}$}. Observing parameters, including beam sizes, channel spacings, etc. are summarized in Table\,\ref{tab-obs}. At a distance of 140\,pc, the $^{12}$CO\,(2-1) mean synthesized FWHM beam size of 0\asp5 corresponds to a linear resolution of 70\,au.

Short-spacing data (maps) for all three lines were obtained in September 2006 with the IRAM 30\,m telescope at Pico Veleta in Spain. They were combined with the interferometric $uv$\ data using the short-spacings processing tool of the {\sc mapping}\footnote{see http://www.iram.fr/IRAMFR/GILDAS} software to help the imaging and deconvolution. The continuum emission was derived from the line-free channels in the interferometric data and subtracted before combining the $uv$\ tables, using the Gildas task {\it uv\_subtract}.
Imaging and deconvolution of the line data was done with natural {\it uv}-weighting and the Hogbom algorithm \citep{hogbom1974}, resulting in the synthesized beam sizes and 1\,$\sigma$\, noise levels in the maps (measured in line-free channels) listed in Table\,\ref{tab-obs}. For the \mbox{$^{12}$CO\,(2--1)} data, we also produced an additional smoothed version at $\approx$1\asp0 resolution by applying $uv$\ tapering with radius 200\,m, as well as a version with robust $uv$\ weighting resulting in an effective angular resolution of $\approx$0\asp45.


\subsection{SMA millimeter observations} \label{sec:obs:sma}

Observations with the Submillimeter Array\footnote{The Submillimeter Array is a joint project between the Smithsonian Astrophysical Observatory and the Academia Sinica Institute of Astronomy and Astrophysics and is funded by the Smithsonian Institution and the Academia Sinica.}  \citep[SMA;][]{ho2004} were made on 2006 December 6 (extended configuration) and December 31 (compact configuration), covering the frequency ranges around 267 and 277\,GHz in the lower and upper sidebands, respectively. This setup includes the lines of \mbox{HCO$^+$(3--2)} (267.558\,GHz) and \mbox{HCN(3--2)} (265.886\,GHz) which were observed with a channel spacing equivalent to 0.23\,\kms.  Typical system temperatures were 350-500\,K. Only the \mbox{HCO$^+$(3--2)} data are used for this paper. The quasar 3C279 was used for bandpass calibration, and the quasars B0355+508 and 3C111 for gain calibration.  Uranus was used for absolute flux calibration, which is accurate to $\approx$20\%. The data were calibrated and imaged using the Miriad toolbox \citep{sault95}. The 1.1\,mm continuum map was constructed using line-free channels in both sidebands. Observing details are summarized in Table \ref{tab-obs}. In addition, Table\,\ref{tab-lines} summarizes the purpose and use of all millimeter molecular lines used in this paper.


\subsection{Optical\,/\,NIR observations observations} \label{sec:obs:nir}

Optical narrow-band images in the H$\alpha$\ (656\,nm) and S[{\sc ii}] (674\,nm) filters, as well as in the $R$-band (641\,nm) were taken with the 2048$\times$2048 CCD prime focus camera (pixel scale 1\farcs23) of the 2\,m Alfred Jensch Telescope\footnote{The diameter of the Schmidt correction plate is 1.34\,m.} at the Thuringian state observatory in Tautenburg/Germany on 2013, August 2. Two exposures were obtained with each filter with integration times of 1200\,s for the narrow-band filters and 180\,s for $R$-band. After the standard image processing, the world coordinate system (WCS) was established using the SCAMP astrometry program \citep{bertin2006}, which yielded an astrometric RMS of 0\farcs15. The new images complement the earlier ones taken in 2001 with the same setup \citep{stecklum04}, thus providing a temporal baseline of about 12 years for the proper motion analysis of HH\,494  (App.\,\ref{sec:app:hh494}).
For compiling the spectral energy distribution (SED; Fig.\,\ref{fig:sed}), we also use photometry from the Spitzer \citep{werner2004} {\it IRAC} \citep{fazio2004} and MIPS \citep{rieke2004} instruments, as well as from the ALLWISE \citep{cutri2013} and unWISE \citep{schlafly2019} catalogs.



\section{Results} \label{sec:res}


\subsection{Overview} \label{ssec:res:ov}

\begin{figure*}[htb]
\begin{center}
\includegraphics[width=1.0\textwidth]{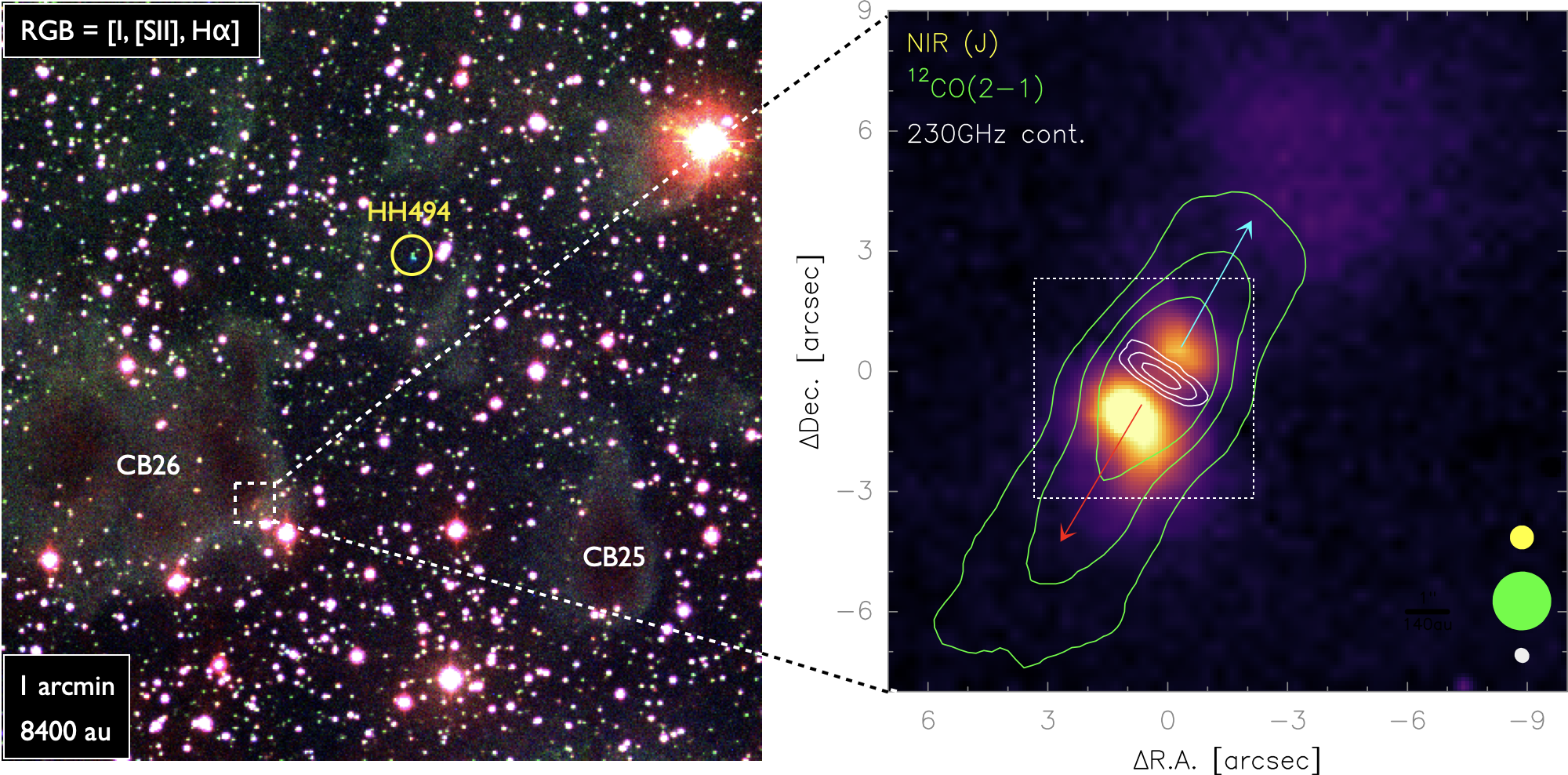}
\caption{\label{fig-overview}
 Overview of the CB\,26 region. The left panel shows a wide-field optical "true-color" image, which is based on H$\alpha$ (blue), [SII] (green), and I--band (red) images \citep{stecklum04}. The globules CB\,25 and CB\,26 as well as Herbig-Haro object HH\,494 are marked in the image. The zoomed-in right panel shows a NIR J-band image of the bipolar reflection nebula (color, 0\asp6 resolution), overlaid with contours of the 1.3\,mm dust continuum emission from the disk (white contours at 3, 7.5, and 15\,mJy/beam). Green contours show the integrated $^{12}$CO(2--1) emission (0.5 to 12\,\kms) from the bipolar molecular outflow as presented in \citet{launhardt09}. The red and blue arrows indicate the large-scale outflow orientation. Beam sizes are shown in respective colors in the lower right corner. The white dashed square marks the image section shown in Fig.\,\ref{fig:contim}. The reference position is  $04^h59^m50.74^s, 52^{\circ}04^{\prime}43.80^{\prime\prime}$\ (J2000).
}
\end{center}
\end{figure*}

\begin{figure}[htb]
\begin{center}
\includegraphics[width=0.48\textwidth]{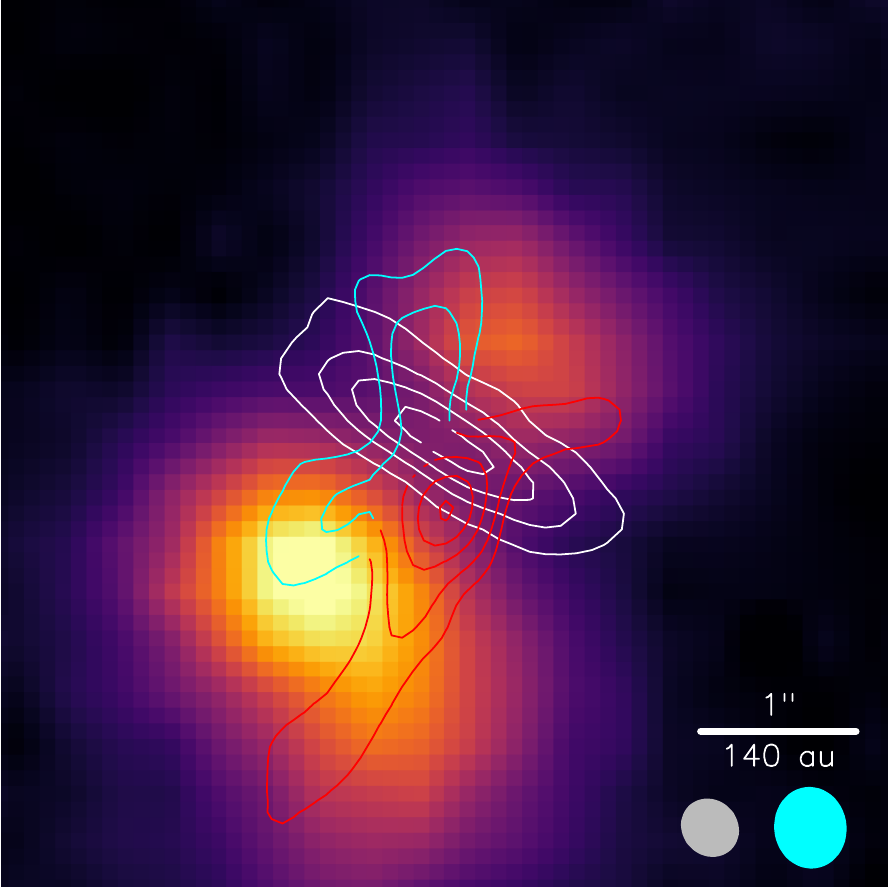}
\caption{\label{fig:contim}
 NIR $J$-band image of the bipolar nebula (see Sect.\,\ref{sec:obs:nir}) overlaid with contours (3, 8, 13, and 18\,mJy/beam) of the 1.3\,mm dust continuum emission from the disk \citep{zhang2021}. Overlaid as blue and red contours, representing the rotation-related blue- and red-shifted sides (cf. Fig.\,\ref{fig_intmaps1}), is the integrated $^{12}$CO(2--1) emission (0.5 to 12\,\kms) from the bipolar molecular outflow as seen in the new, higher-resolution CO(2--1) data, which we present and analyze in this paper. The 1.3\,mm continuum and $^{12}$CO(2--1) synthesized beam sizes as well as the angular and linear scales are indicated in the lower right corner (cf. Fig.\,\ref{fig-overview}).
 }
\end{center}
\end{figure}

Figure\,\ref{fig-overview} shows an overview of the CB\,26 region, with \mbox{CB\,26\,-\,YSO\,1} located at the south eastern tip of the double-core globule CB\,26 and the Herbig-Haro object HH\,494 located at 6\dotmin15 toward the north west \citep[cf.][]{stecklum04}. The optical "true-color" image (left panel) is based on wide-field H$\alpha$ (blue), [SII] (green), and I--band (red) images obtained with the Schmidt CCD camera of the 2-m Alfred Jensch Telescope at the TLS Tautenburg \citep{stecklum04}.

CB\,26\,-\,YSO\,1 is resolved into an edge-on circumstellar disk \citep{launhardt01}, a bipolar NIR reflection nebula with a dark lane at the position of the disk \citep{stecklum04}, and a collimated bipolar molecular outflow that was reported to show kinematic signatures of rotation around its axis with the same orientation as the disk \citep{launhardt09}. The deep $J$-band image of the reflection nebula shown in the right panel of Fig.\,\ref{fig-overview} also shows a faint extended lobe toward the north west, but not toward the south east. Since the north western lobe of the bipolar CO outflow is blue-shifted, and thus slightly directed toward us, this faint extended NIR emission likely marks the region where the north western outflow lobe is breaking out of the dense globule core.

Figure\,\ref{fig:contim} shows the central part of the NIR reflection nebula, again with the 1.3\,mm thermal dust continuum emission overlaid on the central dark lane, but now with the integrated CO emission of the higher-resolution CO(2--1) data, which we present and analyze in this paper. The high-resolution CO(2--1) data show an X-shaped morphology at the location of the disk and the origin of the outflow. The kinematic structure indicates the same corotation of disk and outflow as concluded from the lower-resolution data presented in \citet{launhardt09}, although disk and outflow emission cannot be separated unambiguously close to the disk based on the data only. To separate their contributions to the total emission and enable us to reveal the morphology and kinematics of the outflow as close as possible to its origin, we use the dust continuum emission (Sect.\,\ref{ssec:res:cont}) to establish a physical model of the disk (Sect.\,\ref{ssec:mod:disk}), which we then use to model and subtract the CO emission from the disk.

\begin{figure*}
\includegraphics[width=0.925\textwidth]{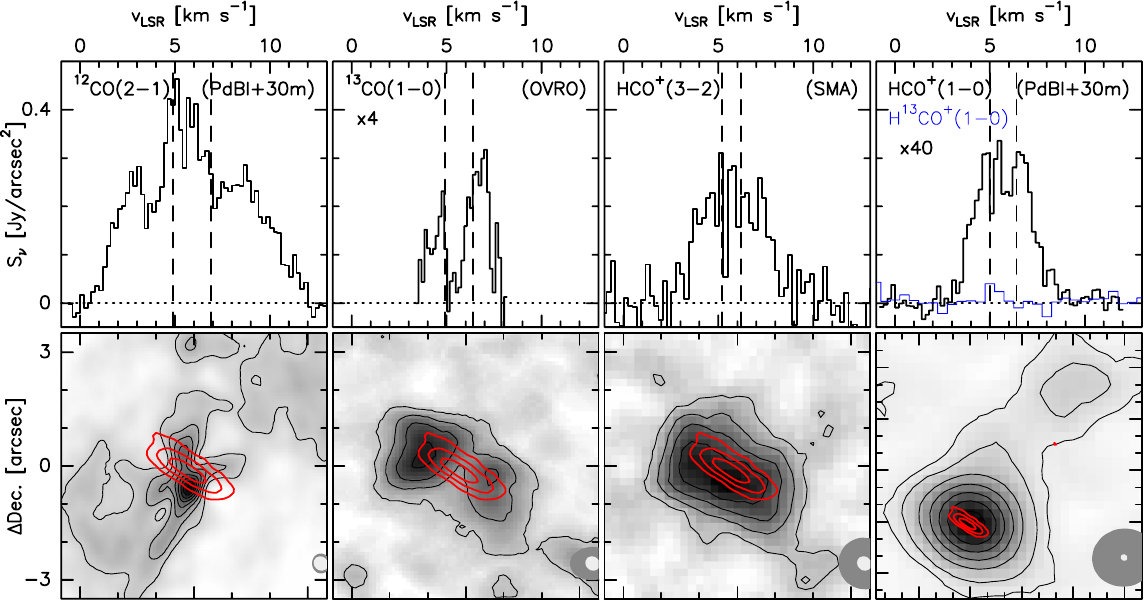}
\includegraphics[width=1.0\textwidth]{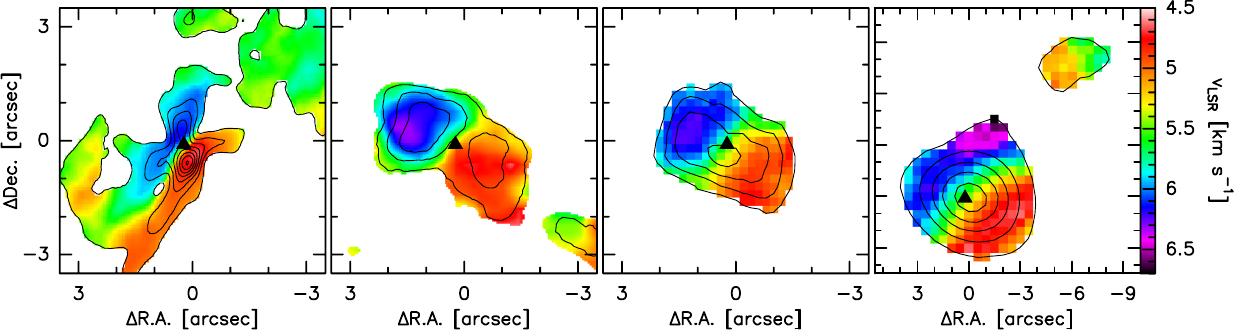}
\caption{\label{fig_intmaps1}
Line spectra of CB\,26, integrated over $\approx$1.5 beams around the disk center (top row), integrated intensity maps (middle row), and mean velocity fields (1$^{\rm st}$\ moment maps, bottom row). From left to right:~~
$^{12}$CO\,(2--1) (0.5 to 4.8 and 7.0 to 12.0\,\kms),
$^{13}$CO\,(1--0) (3.4 to 4.9 and 6.2 to 7.7\,\kms),
HCO$^+$(3--2) (1 to 10\,\kms), and 
HCO$^+$(1--0) and H$^{13}$CO$^+$(1--0) in blue (2.0 to 5.0 and 6.3 to 9.0\,\kms).
Dashed vertical lines in the spectra (top) indicate the velocity range within which the spectra are most strongly affected by resolved-out emission and self-absorption from the extended envelope (see Figs.\,\ref{fig_chanmap_12co_obs} through \ref{fig_chanmap_hco10}). Total intensity contours in the maps start at 3\,$\sigma$. Overlaid as red contours in the middle row at 2.5, 7.5, and 15\,mJy/beam is the 230\,GHz dust continuum emission from the disk. Synthesized FWHM beam sizes are shown as gray ellipses in the lower right corners (dark gray: line, light gray: continuum). The bottom panels show the respective 1$^{\rm st}$\ moment maps with contours of the total intensity overlaid. A black triangle marks the location of the central star (and center of the disk). We note that, except for HCO$^+$\,(3--2), the envelope-dominated central velocity channels were masked out before the moment maps where generated.}
\end{figure*}

Figure\,\ref{fig_intmaps1} shows an overview of the molecular line data used in this paper. The top row shows the spectra of \mbox{$^{12}$CO(2-1)}, \mbox{$^{13}$CO(1-0)}, \mbox{HCO$^+$(1-0)}, and \mbox{HCO$^+$(3-2)} at the position of the disk center (integrated over about 1.5 beams, see Table\,\ref{tab-obs}). The middle row shows the respective integrated intensity maps with contours of the 1.3\,mm dust continuum emission from the disk overlaid. The bottom row shows the $1^{st}$\ moment maps of all four lines. The corresponding channel maps are shown in Figs.\,\ref{fig_chanmap_12co_obs} and \ref{fig_chanmap_13co} through \ref{fig_chanmap_hco32}. The H$^{13}$CO$^+$(1-0) line was only marginally detected when integrating over the entire primary beam area; a channel map is therefore not shown here.

The integrated intensity maps of $^{13}$CO(1-0) and HCO$^+$(3-2) show that we have mostly detected and spatially resolved the emission from the disk (Fig.\,\ref{fig_intmaps1}). The HCO$^+$(1-0) map is also dominated by emission from the position of the disk, but the disk is not resolved spatially and some emission from the north west lobe of the outflow (and the envelope) is also recovered. In contrast, the $^{12}$CO(2-1) map shows a very different morphology. It resembles an ``X'' or butterfly, centered on the disk, but with no emission detected from about the outer half of the disk (midplane). In Sects.\,\ref{ssec:res:12co} and \ref{sec:mod}, we discuss and model this peculiar morphology in terms of a flared disk with a resolved-out (and possibly self-absorbing) envelope and a collimated disk wind seen edge-on. The 1$^{st}$\ moment maps, also shown in Fig.\,\ref{fig_intmaps1} (bottom panels), depict clearly the Keplerian rotation of both the disk and the molecular disk wind, the latter one in $^{12}$CO(2-1) only. 
Since \mbox{$^{12}$CO(2-1)} is the only line that traces the outflow, and the main purpose of this study is to reveal the nature and morphology of the outflow, we analyze quantitatively only the \mbox{$^{12}$CO(2-1)} data, and use the other lines only for qualitative arguments on the disk rotation.

The position angle (P.A.) of the disk and outflow axis was derived from the three continuum images \citep{zhang2021} and the low-resolution $^{12}$CO(2-1) map (Fig.\,\ref{fig_covelfield}) to be P.A.\,=\,$148\pm 1\degr$\ (E of N). This value agrees remarkably well with the relative P.A.\,=\,$147.5\pm 0.5\degr$\ (with regard to CB\,26-YSO\,1) of the Herbig-Haro object HH\,494 at a projected separation of 6.15\arcmin\ (corresponding to $\approx 5\times10^4$\,au or 0.25\,pc at a distance of 140\,pc) reported by \citet[][see also Fig.\,\ref{fig-overview} and App.\,\ref{sec:app:hh494}]{stecklum04}.

In Section\,\ref{sec:mod}, we analyze, model, and discuss these data in terms of morphology and kinematics of both the disk and the outflow, and their relation. For this purpose, we rotate all maps counterclockwise by $32\degr$\ (such that P.A.$^{\prime}$\,=\,$180\degr$) in order to align the disk and outflow axes with the $z$-axis. We furthermore subtract the systemic velocity $v_0=5.95$\,\kms\ (Sect.\,\ref{ssec:res:12co}) from the line spectra and refer in the following to $\Delta v = v_\mathrm{LSR} - v_0$.


\subsection{Dust continuum emission} 
\label{ssec:res:cont}

\begin{figure}[htb]
\begin{center}
\includegraphics[width=0.48\textwidth]{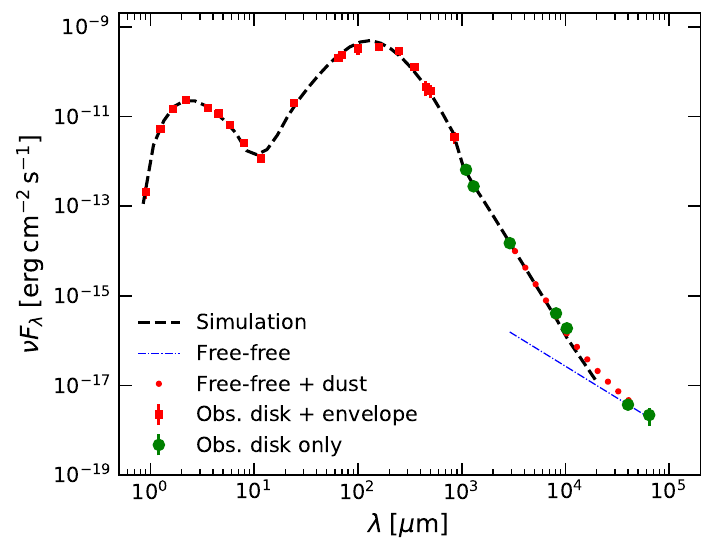}
\caption{\label{fig:sed}
Spectral energy distribution of CB\,26-YSO\,1, adopted from \citet{zhang2021}, where we also list the fluxes in Table\,1. Red squares and green dots with error bars show the observational data. The dashed black line shows best-fit model SED of \citet{zhang2021}. The dashed-dotted blue line represents the adopted contribution from free-free emission. The dotted red line corresponds to the sum flux of dust and free-free emission between 2.9\,mm and 4.0\,cm. The break in the SED at $\sim10^3\,\mu$m is due to the transition from total flux measurements at shorter wavelengths to interferometric data at longer wavelengths, which do not recover the total flux from the extended envelope emission.
}
\end{center}
\end{figure}

\begin{table}[ht]
\caption{Geometrical disk parameters derived in other studies.}
\label{tab:diskpar} 
\begin{tabular}{llll}
\hline \hline
Paper & $R_{\rm in}^{\rm dust}$\ [au] & $R_{\rm out}^{\rm dust}$\ [au] & $i_{\rm disk}$\ [\adeg] \\
\hline 
\citet{launhardt09} & $\ldots$         & $\ldots$          & $85\pm4$ \\[1.0mm]
\citet{sauter09}    & $45\pm5$         & $200\pm25$        & $85\pm5$ \\[1.0mm]
\citet{akimkin12}   & $37^{+16}_{-14}$ & $222^{+75}_{-57}$ & $82^{+3}_{-17}$ \\[1.0mm]
\citet{zhang2021}   & $16^{+37}_{-8}$  & $172^{+20}_{-23}$ & $88^{+2}_{-5}$ \\[1.0mm]
\hline
\end{tabular} 
\end{table}

Early single-dish observations of CB\,26 revealed a compact and basically unresolved 1.3\,mm dust continuum source with a total flux density of $S_{\nu}^{\rm 1.3mm}=240\pm50$\,mJy \citep{launhardt97,launhardt10}. Subsequent interferometric observations showed that $\approx$50\% of this emission originates from an edge-on protoplanetary disk with an outer radius of $r_{\rm out}=172\pm22$\,au, an inner hole of size  $r_{\rm in}=16^{+37}_{-8}$\,au, and a total (dust+gas) mass of $M_{\rm disk}\approx0.1$\msol\ \citep{launhardt01,zhang2021}. The other 50\% of the 1.3\,mm dust continuum emission was attributed to a remnant envelope (or remnant dense globule core) of poorly constrained size ($\approx10^4$\,au) and total mass of $M_{\rm env}\approx$\,0.1--0.2\,\msol\ \citep{launhardt10,launhardt13,zhang2021}. Figure\,\ref{fig:contim} shows our highest-resolution dust continuum image of the CB\,26 disk, obtained by combining interferometric 230\,GHz continuum data from PdBI and OVRO \citep{zhang2021}, overlaid on the NIR $J$-band image of the bipolar nebula. 

The high-resolution dust continuum images of the CB\,26 disk at $\lambda=1.1$\,mm ($\nu=268$\,GHz), 1.3\,mm (230\,GHz), and 2.9\,mm (102\,GHz), together with the SED (Fig.\,\ref{fig:sed}) were already used by \citet{sauter09}, \citet{akimkin12}, and \citet{zhang2021} to derive a physical model of the disk by means of continuum radiative transfer modeling. All three studies, albeit using slightly different approaches and slightly different versions of the data, derive very similar basic parameters for the CB\,26 disk, which we summarize in Table\,\ref{tab:diskpar}. The values for $R_{\rm in}^{\rm dust}$\ may look very different at first glance, but their error bars fully overlap with all three estimates predicting an upper limit of $\approx$50\,au.

In Sect.\,\ref{sec:mod}, we use a physical model of the disk with these parameters, which are summarized in Table\,\ref{tab-diskpar}, to derive a model of the $^{12}$CO(2--1) emission from the disk with the two-fold aim of {\it (i)} further constraining certain parameters of the disk (e.g., its inclination) and the central star (e.g., its kinematic mass), and {\it (ii)} subtracting this model from the CO data in order to better reveal the morphology and kinematics of the disk wind at its launch region on the disk.


\subsection{$^{12}$CO(2--1)} \label{ssec:res:12co}

\begin{figure*}[htb]
\begin{center}
\includegraphics[width=1.0\textwidth]{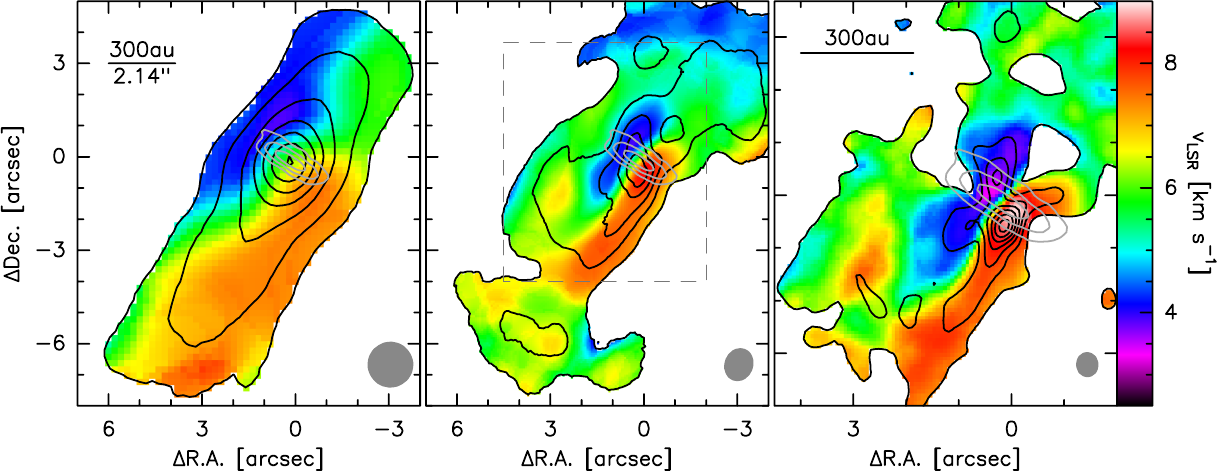}
\caption{\label{fig_covelfield}
 $^{12}$CO(2--1) mean velocity field (1$^{st}$\ moment map, color) and integrated intensity (black contours) of CB\,26 at successively higher angular resolution.
 Left: as presented in \citet{launhardt09} at 1\asp47 resolution.
 Middle: with the complete data set presented in this paper, but mapped with a $uv$\ taper of radius 200\,m and natural weighting (to restore a maximum of extended emission), resulting in an effective angular resolution of $\approx$1\asp0. 
 Right: same as middle, but mapped without tapering and with robust $uv$\ weighting, resulting in an effective angular resolution of $\approx$0\asp45. 
 This image shows a smaller (zoomed) area than the other two maps, as indicated by the gray dashed rectangle in the middle panel. Synthesized FWHM beam sizes are shown as gray ellipses in the lower right corners of each panel. Overlaid as gray contours at 2.5, 7.5, and 15\,mJy/beam is the 230\,GHz dust continuum emission from the disk.). 
}
\end{center}
\end{figure*}

The \mbox{$^{12}$CO(2--1)} channel maps are shown in Fig.\,\ref{fig_chanmap_12co_obs} and the spectrum at the position of the disk center is shown in Fig.\,\ref{fig_intmaps1}. CO emission is detected in the entire velocity range \mbox{$\Delta v \approx -5.2$\ to +5.9\,\kms}, translating into a Keplerian radius range of $r<-18$\,au and $>$+14\,\au, respectively. However, we do not infer the size of a possible "CO hole" directly from this lack of emission at larger relative velocities, but refer to the modeling of the CO emission in Sect.\,\ref{ssec:mod:disk}.
The outer velocity channels in the range \mbox{$|\Delta v| \approx 4\ldots6$\,\kms} show relatively compact emission, slightly shifted toward the north west and south east, respectively, with regard to the center of the disk (continuum image). This emission is likely to represent the inner, fast rotating part of the Keplerian disk (cf. Fig.\,\ref{fig_chanmap_12co_mod}).
At lower relative velocities, in the range \mbox{|$\Delta v| \approx 2.4\ldots3.4$\,\kms}, narrow streaks extending perpendicular to the plane of the disk become evident. This is likely to be the signature of a disk wind, the morphology and kinematics of which we analyze in more detail in Sect.\,\ref{sec:mod}.
At even lower relative velocities, in the range \mbox{|$\Delta v| \approx 1.4\ldots2.2$\,\kms}, these streaks become wider and are complemented by a strong and asymmetric butterfly-like structure in the middle, with left (north east) and right (south west) parts strictly separated between lower and higher velocities. We show below (Sect.\,\ref{sec:mod}) that this butterfly structure originates from the warm atmosphere of the flared Keplerian disk, while the streaky extensions perpendicular to the disk originate from a rotating disk wind. The features described above are more clearly visualized in the binned channel maps of the observed CO emission shown in Fig.\,\ref{fig_binchan} (top panels).

At the lowest relative velocities, in the range \mbox{$|\Delta v| \lesssim 1.2$\,\kms}, the structure of the recovered \mbox{$^{12}$CO(2--1)} emission becomes very complex and fainter. Despite the complementation by short-spacing data from single-dish observations, these central velocity channels could not be restored well and suffer from resolved-out extended emission and possibly also from self-absorption. They were therefore masked for generating the moment maps (Figs.\,\ref{fig_intmaps1} and \ref{fig_covelfield}) and are not considered in the subsequent modeling and analysis.

At an effective angular resolution of $<0.5$\arcsec, the total intensity map (0$^{th}$\ moment) of \mbox{$^{12}$CO(2--1)} (Figs.\,\ref{fig_intmaps1} and \ref{fig_covelfield}) shows an X-shaped morphology with emission coming from both the disk itself and an outflow or disk wind. The 1$^{st}$\ moment map shown in the same figures indicates that the entire X-shaped CO structure, including the extended vertical streaks, is coherently rotating with the same orientation as the disk, as illustrated by the 1$^{st}$\ moment maps of $^{13}$CO(2--1), HCO$^+$(1--0) and (3--2), which are also shown in Fig.\,\ref{fig_intmaps1}.
In order to illustrate how this inner resolved X-shaped structure around the disk is related to the image of the smooth rotating molecular outflow presented in \citet{launhardt09}, we also show in Fig.\,\ref{fig_covelfield} the original 1$^{st}$\ moment map from 2009, as well as a smoothed image produced from the new data with $uv$\ tapering (Gaussian radius 200\,m) resulting in $\approx$\,1\asp0 resolution. The new, high-resolution image, which reveals the inner X-shaped structure with the rotation signature, is fully compatible  with the smooth image of the extended rotating outflow presented in \citet{launhardt09}. Not surprisingly, the addition of the longer-baseline data did not significantly affect the flux and the new high-resolution CO map recovers the same total CO flux as the lower-resolution map from 2009 (35$\pm$3\,Jy\,\kms). We analyze the flux distribution in more detail in Sect.\,\ref{ssec:mod:outflow1}.

\begin{figure}
\begin{center}
\includegraphics[width=0.48\textwidth]{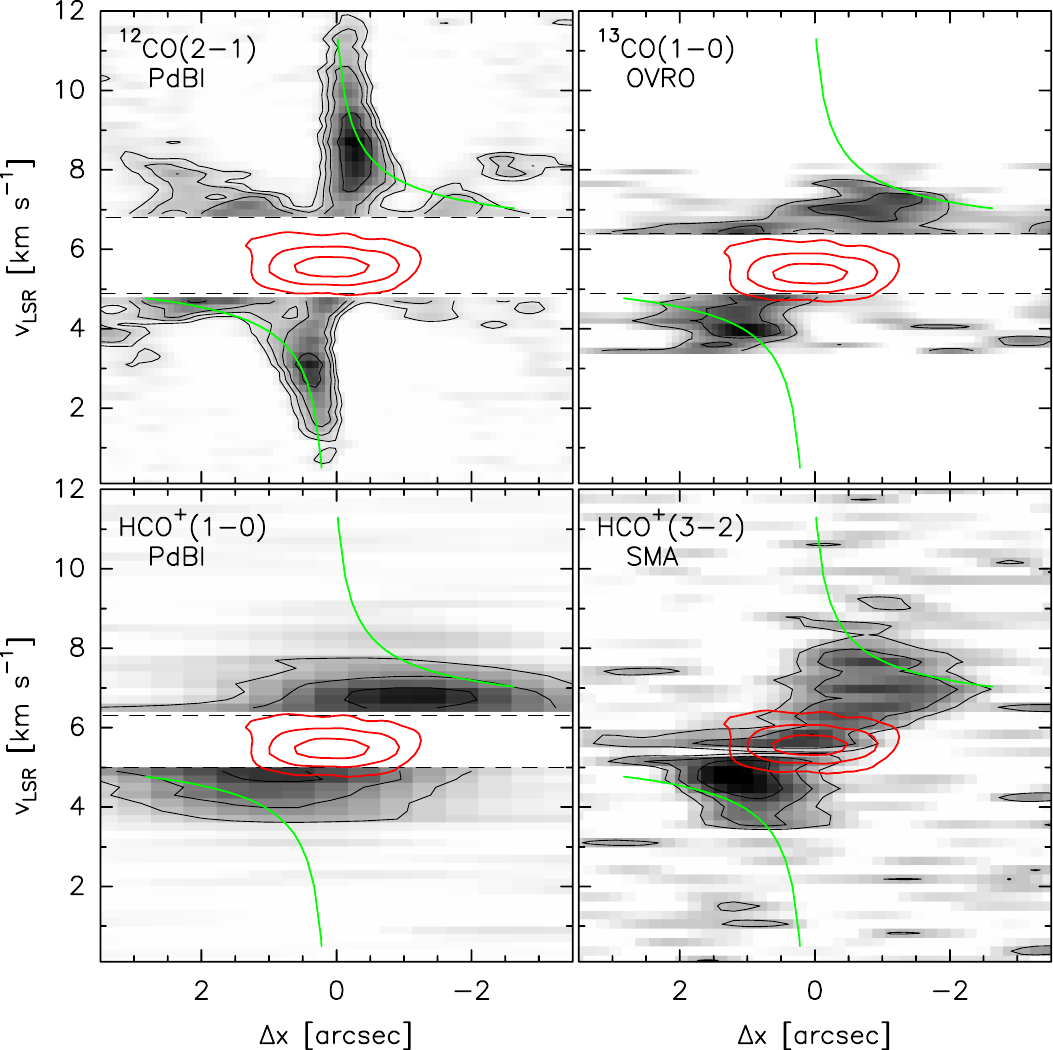}
\caption{\label{fig_pvdisk}
 Position-velocity diagrams along the plane of the disk of $^{12}$CO\,(2--1), $^{13}$CO\,(1--0), HCO$^+$(1--0), and HCO$^+$(3--2). Contours start at 3\,$\sigma$. Except for HCO$^+$(3--2), the envelope-dominated central velocity channels were masked out for the analysis (see also Figs.\,\ref{fig_chanmap_12co_obs}, \ref{fig_chanmap_13co} and \ref{fig_chanmap_hco10}). Overlaid as red contours for comparison (with the y-axis being $\Delta$y instead of $v_{\rm LSR}$) is the 230\,GHz dust continuum emission from the disk (2.5, 7.5, and 15\,mJy/beam), rotated by 32$\degr$. Green lines show the Keplerian rotation curve for a central mass of 0.55\,\msol\ and systemic velocity of $v_0 = 5.95$\,\kms\ (only to guide the eye; this is not the fit to the data).
}
\end{center}
\end{figure}

Figure\,\ref{fig_pvdisk} shows the position-velocity diagram (PVD) of \mbox{$^{12}$CO(2--1)} along the mid-plane of the disk, along with PVDs of the other three lines. The Keplerian rotation velocity field can be clearly seen in the \mbox{$^{12}$CO} PVD, with the drawback that the emission from the outer parts of the disk at lower relative velocities might be affected the velocity overlap with the surrounding self-absorbing envelope, although the Keplerian velocity at the outer edge of the disk ($R^{\rm dust}_{\rm out}\approx200$\,au) would still be $v_{\rm Kep}\approx1.6$\,\kms, which is significantly larger than the affected $\Delta v\approx\pm1.2$\,\kms.
From the symmetry of the \mbox{$^{12}$CO(2--1)} (and \mbox{HCO(3--2)}) PVDs (mirrored and cross-correlated), we derive a systemic velocity for the CB\,26 disk of $v_0=5.95\pm0.05$\,\kms.
The \mbox{$^{12}$CO(2--1)} velocity field can be fit with a Keplerian disk around a central mass of $M_{\ast}=0.55\pm0.1$\,\msol\ (Sect.\,\ref{sec:mod}). To guide the eye in the PVDs (Fig.\,\ref{fig_pvdisk}), we overplot a Keplerian rotation curve for $M_{\ast}=0.55$\,\msol\ and \mbox{$v_0 = 5.95$\,\kms}; this is not a fit to the data. A more detailed analysis of the velocity field, including an attempt to separate the contribution from the disk and the outflow (disk wind) is undertaken in Sect.\,\ref{sec:mod}.


\subsection{$^{13}$CO(1--0)} \label{ssec:res:13co}

The \mbox{$^{13}$CO(1--0)} channel maps are shown in Fig.\,\ref{fig_chanmap_13co} and the spectrum at the position of the disk center is shown in Fig.\,\ref{fig_intmaps1}. Due to the narrow spectral bandwidth of the correlator (4.8\,\kms, Sect.\,\ref{sec:obs:ovro}), only the relative velocity range between -2.5\,\kms\ and +2.1\,\kms\ is covered, translating into a Keplerian radius range of $r<-80$\,au and $>$110\,\au, respectively, which means that the observations do not trace the inner $|r|\approx\pm$100\,au of the disk, even if there would be \mbox{$^{13}$CO(1--0)} emission from this range. Emission from the disk is detected in the relative velocity range from \mbox{$\Delta v\approx-2.35$\ to 1.65\,\kms}, translating into a Keplerian radius range of $r<-90$\,au and $>$180\,\au, respectively.
The central velocity channels between \mbox{$\Delta v\approx-1.0$\ and 0.5\,\kms} could not be restored well and also suffer from resolved-out extended emission (plus possibly self-absorption). Similar to \mbox{$^{12}$CO(2--1)}, these channels were therefore masked for generating the moment maps (Fig.\ref{fig_intmaps1}).
The envelope and the outflow are not, or only marginally, detected at low relative velocities, and may also suffer from resolved-out structure and self-absorption. The total intensity map (0$^{th}$\ moment) of \mbox{$^{13}$CO(1--0)} (Fig.\ref{fig_intmaps1}) traces the disk with most of the emission coming from the outer regions. 
The 1$^{st}$\ moment map, also shown in Fig.\,\ref{fig_intmaps1}, indicates the same rotation signature as the other lines, indicative of a Keplerian disk. This can also be seen in the PVD shown in Fig.\,\ref{fig_pvdisk}.


\subsection{HCO$^+$} \label{ssec:res:hco}

The channel maps of \mbox{HCO$^+$(1--0)} and \mbox{HCO$^+$(3--2)} are shown in Figs.\,\ref{fig_chanmap_hco10} and \ref{fig_chanmap_hco32} and the spectra at the position of the disk center are shown in Fig.\,\ref{fig_intmaps1}.
\mbox{HCO$^+$(1--0)} emission from the disk is detected in the relative velocity range \mbox{$|\Delta v|\approx\pm2.2$,\kms}, translating into a Keplerian radius range of $|r|>80$\,au. Some extended emission mainly from the north west outflow lobe (blue, pointing toward us) is also detected in the relative velocity range $-2.2$\ to $-1$\,\kms. Despite the complementation by short-spacing data from single-dish observations, the central velocity channels of \mbox{HCO$^+$(1--0)} are corrupted between $\Delta v\approx-0.8$\ and 0.4\,\kms\ by resolved-out emission and self-absorption from the extended envelope. Similar to \mbox{$^{12}$CO(2--1)}, these channels were therefore masked before generating the moment maps (Fig.\,\ref{fig_intmaps1}). The total intensity map of \mbox{HCO$^+$(1--0)} (Fig.\,\ref{fig_intmaps1}) shows the disk basically unresolved plus some extended emission from the envelope and the north west outflow lobe.
\mbox{H$^{13}$CO$^+$(1--0)} emission is only marginally detected in the relative velocity range $-1$\ to 0.1\,\kms\ and only when integrating over the entire primary beam area, indicating that it most likely originates in the envelope.

\mbox{HCO$^+$(3--2)} emission is also detected in the relative velocity range \mbox{$|\Delta v|\approx\pm2.2$\,\kms}, translating into a Keplerian radius range of $|r|>80$\,au. The central velocity channels between \mbox{$\Delta v\approx -0.7$} and $0.3$\,\kms\ might also be slightly affected by resolved-out emission and self-absorption from the extended envelope (see also Fig.\,\ref{fig_intmaps1}). Since this effect is much smaller than in the lower-excited line, we did not mask these channels. In contrast to \mbox{HCO$^+$(1--0)}, the total intensity map of \mbox{HCO$^+$(3--2)} shows the disk slightly resolved and no traces of the envelope or outflow. The 1$^{st}$\ moment maps of both HCO$^+$\ lines (Fig.\,\ref{fig_intmaps1}), as well as the PVDs along the plane of the disk (Fig.\,\ref{fig_pvdisk}) indicate the same rotation signatures as the CO lines.


\section{Modeling and analysis} \label{sec:mod}

\subsection{Disk model} \label{ssec:mod:disk}

A physical model of the disk was generated with the DUSK code, which described in detail in \citet{akimkin12}. The DUSK code calculates axially symmetric density and temperature distributions assuming that the disk is Keplerian and remains in vertical hydrostatic and thermal equilibrium. It is heated mostly by stellar radiation and is cooled by its own infrared emission. Dust and gas are assumed to be well-mixed and have the same temperature. The main parameters of the disk model are
{\it (i)} the inner and outer radius of the disk, $R_{\rm in}$\ and $R_{\rm out}$,
{\it (ii)} the gas surface density distribution of the disk, $\Sigma^{\rm gas}(r)$,
{\it (iii)} the inclination of the disk, $i_{\rm disk}$, and
{\it (iii)} the mass, $M_{\ast}$, effective temperature $T_{\rm eff,\ast}$, and luminosity, $L_{\ast}$, of the central star.

Parameters $R_{\rm out}$, $\Sigma^{\rm gas}(r)$\ are taken from \citet{akimkin12} from their best-fit model of CB\,26 "230+270 GHz" (see their Table\,2). The fraction of stellar radiation that is intercepted and absorbed by the disk is derived by \citet{akimkin12} to be 0.1\,\lsol. Under the assumption of a razor-thin infinitive disk, which intercepts one quarter of the stellar radiation \citep{armitage2010}, the bolometric luminosity of the system (stellar luminosity) should then be at least 0.4\,\lsol. However, in the model of \citet{akimkin12}, the stellar luminosity degenerates with disk surface density (see their Fig.7), so that the derived fraction 0.1\,\lsol\ is uncertain within a factor of few. In our current modeling, we do not vary the stellar luminosity and disk surface density to keep the number of parameters manageable.

The total mass of the adopted disk model is $\approx$0.2\,\msol\ when assuming a gas-to-dust mass ratio of 100. The mass of the central star(s) will be constrained from the Keplerian velocity field of the disk and is therefore a free parameter in the disk modeling. The same applies to the disk inclination, which is only loosely constrained by the continuum data. The size of the dust emission-free inner hole of $R_{\rm in}^{\rm dust}\approx 37\pm15$\,au derived by \citet{akimkin12} \citep[see also][]{sauter09} is not compatible with the observed velocity structure of \mbox{$^{12}$CO(2--1)}. In Sect.\,\ref{ssec:res:12co}, we show that CO emission is clearly detected from rotational velocities up to $\approx5.9$\,\kms, which corresponds to an inner Keplerian radius of $R_{\rm in}^{\rm gas}\approx 14\pm4$\,au, given the range of possible stellar masses that fit the Keplerian velocity field. It is therefore safe to assume that the dust hole in the disk is not gas-free and the gas disk extends down to an inner radius of about 10\,au or smaller. On the other hand, \citet{zhang2021}, in their recent analysis, show that the dust hole could actually also be as small as 10\,au ($16_{-8}^{+37}$\,au). Therefore, we also treat $R_{\rm in}^{\rm gas}$\ as a free parameter in the disk modeling.

The consistency between modeled and observed \mbox{CO(2--1)} emission was measured with the following (reduced) $\chi^{2}$-criterion:
\begin{equation}
\chi^2 = \frac{1}{N \sigma^2}\sum \limits_{i=1}^{N} \left(T_{i}^{\rm mod} - T_{i}^{\rm obs}\right)^{2},
\label{eq:chi2}
\end{equation}
where $T_{i}^\mathrm{obs}$\ and $T_{i}^\mathrm{mod}$\ are the observed and modeled intensities at a given spatial and spectral pixel, $N$\ is the number of adopted data points, and $\sigma$\ is the noise level in the observed spectra. The sum is calculated only over those spatial and spectral pixels which satisfy the following two conditions. To minimize contributions from the outflow in the observed spectra, we compare model and observations only at those spatial positions where the model intensity integrated over the line profile is higher than a certain threshold, which we chose to be 1\,K$\cdot$\kms. We also exclude the velocity channels at $\Delta v =\pm1.2$\,\kms\ around the central velocity since they are resolved-out by the interferometer and may be affected by self-absorption from the envelope (see Sect.\,\ref{ssec:res:12co}). With these criteria, we effectively use $\approx$50000 out of the $102\times 102\times 61 = 634644$\ total data points in the observed data cube.

\begin{figure}[htb]
\begin{center}
\includegraphics[width=0.48\textwidth]{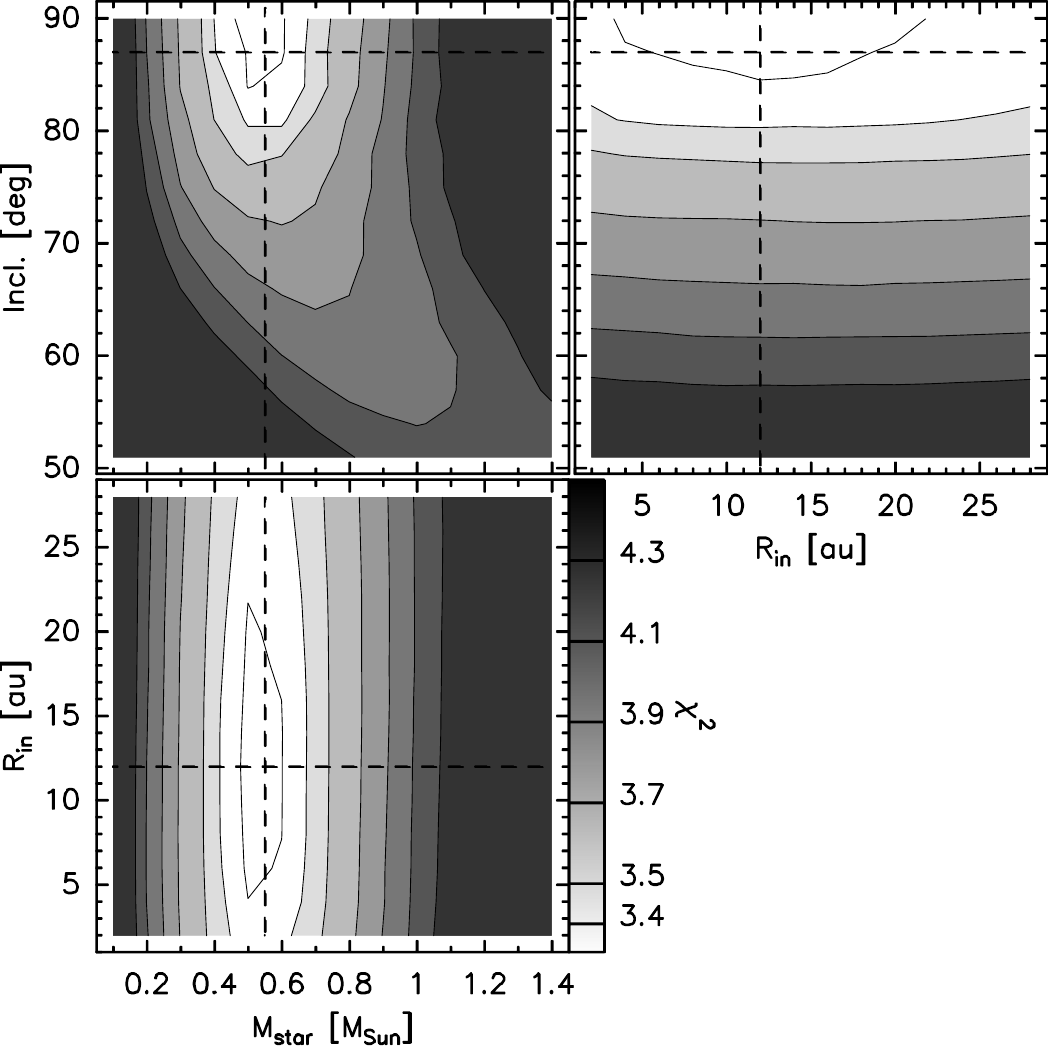}
\caption{\label{fig-mod-chi2}
Marginal two-parameter $\chi^2$\ maps for model grid parameters $M_{\ast}$, $i_{\rm disk}$, and $R_{\rm in}^{\rm gas}$\ derived by comparing the synthetic maps of the \mbox{$^{12}$CO(2--1)} disk models with the observations (robust deconvolution). Dashed lines indicate the parameter values for the chosen "best-fit" disk model that was then subtracted from the data.}
\end{center}
\end{figure}

\begin{figure}[htb]
\begin{center}
%
\includegraphics[width=0.40\textwidth]{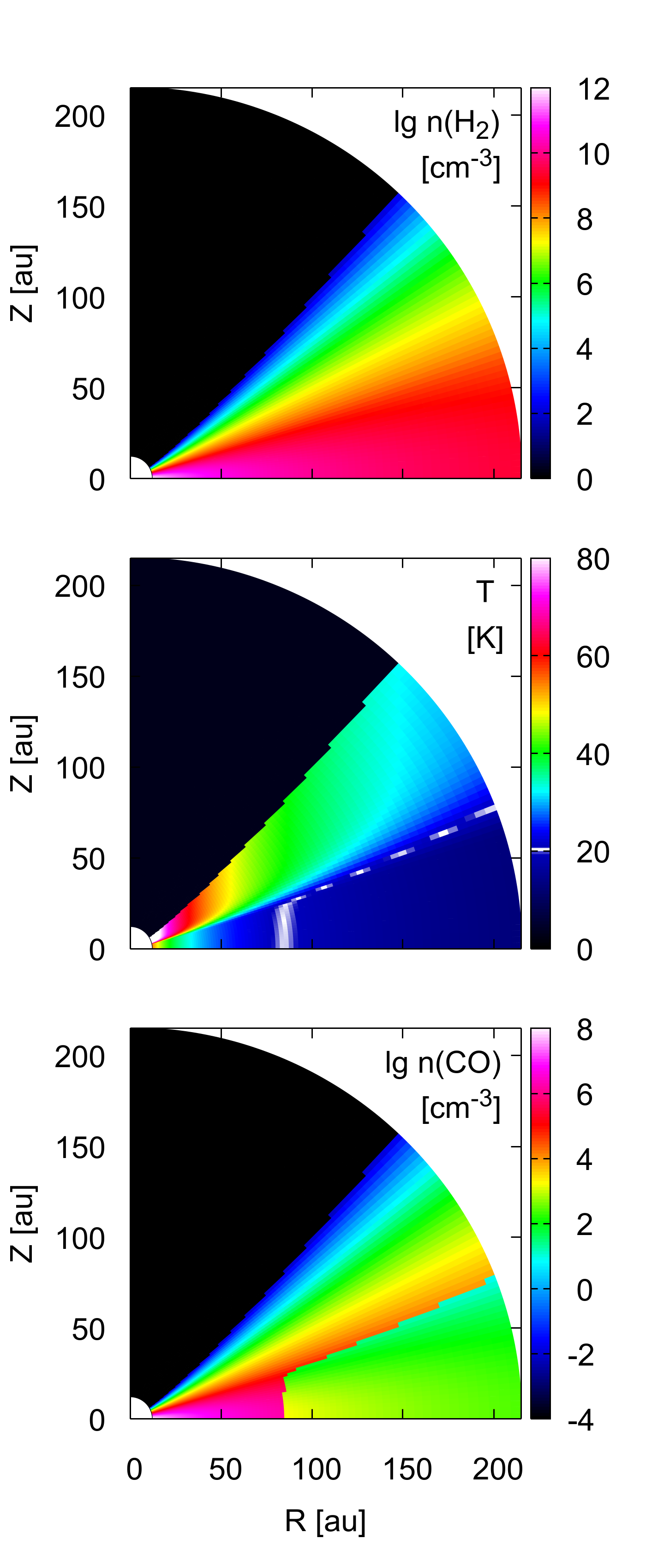}
\caption{\label{fig-mod-disk}
Hydrogen volume density, kinetic gas temperature, and CO volume density distribution of the best-fit disk model that was used to calculate the synthetic CO(2--1) channel maps of the disk emission (Fig.\,\ref{fig_chanmap_12co_mod}). The white line in the middle panel marks the CO snow line at $T_{\rm gas}=20$\,K.
}
\end{center}
\end{figure}

\begin{table}[ht]
\caption{Parameters of the best-fit disk CO model}
\label{tab-diskpar} 
{\footnotesize  
\begin{tabular}{lll}
\hline \hline
Parameter & Value & Remark/reference  \\
\hline 
$R_{\rm in}^{\rm gas}$\ [au]      & 12       & CO modeling, grid range $2\ldots30$\\
$R_{\rm out}$\ [au]               & 220      & \citet{akimkin12} \\
$\Sigma_0^{\rm gas}$\ [g\,cm$^2$] & 710\tablefootmark{a}      & \citet{akimkin12} \\
$p$                               & $-0.81$\tablefootmark{a}  & \citet{akimkin12} \\
$T_{\rm eff,\ast}$\ [K]           & 3700     &  see Sect.\,\ref{ssec:dis:disk}  \\
$L_{\ast}$\ [\lsol]               & 1.0      & see  App.\,\ref{sec:app:lstar}  \\
$i_{\rm disk}$\ [deg]             & 87       & CO modeling, grid range $45\ldots90$\\
$M_{\ast}$\ [\msol]               & 0.55     & CO modeling, grid range $0.1\ldots1.4$\\
\hline
\end{tabular} \tablefoot{
\tablefoottext{a}{$\Sigma^{\rm gas}=\Sigma_0^{\rm gas}\,\left(\frac{r}{1\,{\rm AU}}\right)^p$}
}}
\end{table}

The star mass, disk inclination, and inner radius of the gas disk are the only free parameters for our initial grid of models. Parameter ranges of the model grid are listed in Table\,\ref{tab-diskpar}. The synthetic molecular channel maps are calculated with the URAN(IA) code described in \citet{pavlyuchenkov07}. The abundance of CO is described as: $X(CO)=1\times10^{-4}$ where $T>20$\,K and $X(CO)=1\times10^{-7}$\ where if $T\le20$\,K, to account for the CO snowline \citep[e.g.,][]{bosman2018}. The synthetic maps are convolved with the respective Gaussian clean beam before they are compared to the observed \mbox{$^{12}$CO(2--1)} maps derived with robust deconvolution (see Table\,\ref{tab-obs}). 
Figure\,\ref{fig-mod-chi2} shows the marginal two-parameter $\chi^2$\ maps for the three model grid parameters, from which we determine the best-fit model that we use to subtract from the data. For the disk inclination, we derive a value of $i_{\rm disk}=87\pm2$\degr, consistent with earlier estimates (see Table\,\ref{tab:diskpar}). For the size of the inner hole in CO, we derive a value of $R_{\rm in}^{\rm gas}=12\pm6$\,au. For the mass of the central star, we derive $M_{\ast}=0.55\pm0.1$\,\msol. Table\,\ref{tab-diskpar} summarizes all parameters of the best-fit disk model that we use in this paper, including the parameters adopted from \citet{akimkin12} and the ranges and best-fit values of the parameters evaluated here in the CO modeling. Figure\,\ref{fig-mod-disk} shows the distributions of the hydrogen and CO volume density and kinetic gas temperature of the best-fit disk model that was used to calculate the synthetic CO(2--1) channel maps of the disk emission. Clearly visible in the middle panel is also the CO iceline at $\sim$80\,au, where the mid-plane temperature drops below 20\,K.

The \mbox{$^{12}$CO(2--1)} channel maps of the best-fit disk model are shown in Fig.\,\ref{fig_chanmap_12co_mod}. These channel maps are then subtracted from the observed channel maps (Fig.\,\ref{fig_chanmap_12co_obs}). The residual \mbox{$^{12}$CO(2--1)} emission (Fig.\,\ref{fig_chanmap_12co_diff}) is now considered a much better representation of the disk wind than the total observed emission, albeit we still have to consider possible residual disk contamination due to model imperfections and the simple subtraction procedure. It nevertheless allows us to trace the outflow closer to the disk than with the combined emission. The subsequent analysis of the outflow morphology is therefore based on these residual channel maps (Fig.\,\ref{fig_chanmap_12co_diff}).


\subsection{Outflow analysis} \label{ssec:mod:outflow1}

\begin{figure}
\begin{center}
\includegraphics[width=0.48\textwidth]{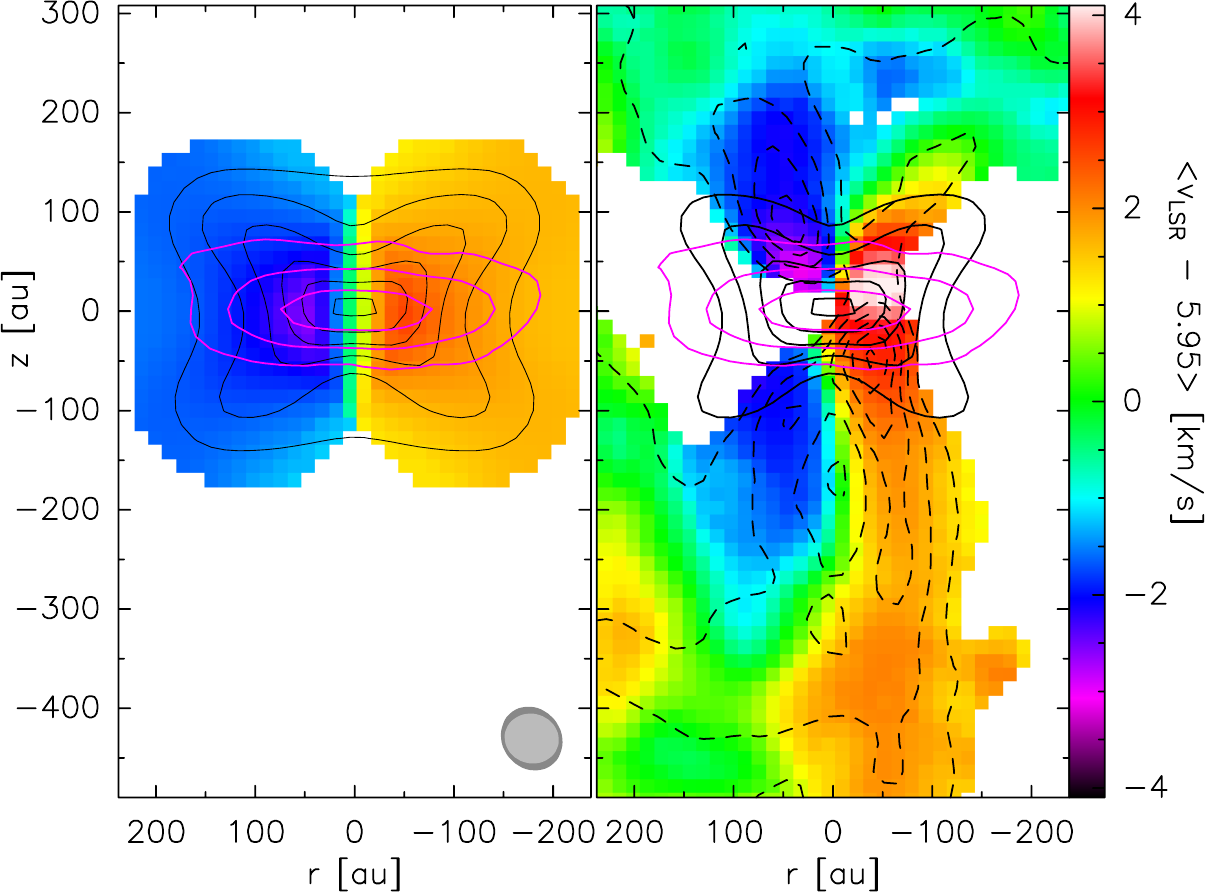}
\caption{\label{fig_velfield}
 Mean velocity field ($1^{\mathrm st}$\ moment maps) of $^{12}$CO\,(2--1) in the velocity range $\pm6$\,\kms\ of the best-fit disk model (left; Table\,\ref{tab-diskpar} and Fig.\,\ref{fig_chanmap_12co_mod}) and the residual CO emission after subtracting the disk model and masking the corrupted central velocity channels between $\pm1.2$\,\kms\ (right). Black solid contours show the total integrated intensity of the CO emission from the model disk. Magenta contours show the 230\,GHz dust continuum emission from the disk at 2.5, 7.5, and 15\,mJy/beam. Dashed contours show the total integrated intensity of the residual CO emission. Synthesized FWHM beam sizes are shown as gray ellipses in the lower right corner (dark gray: CO, light gray: dust). 
 The maps are rotated counterclockwise by 32$^{\circ}$\ such that the disk and outflow axes are aligned with the z-axis. 
}
\end{center}
\end{figure}

\begin{figure}
\begin{center}
\includegraphics[width=0.48\textwidth]{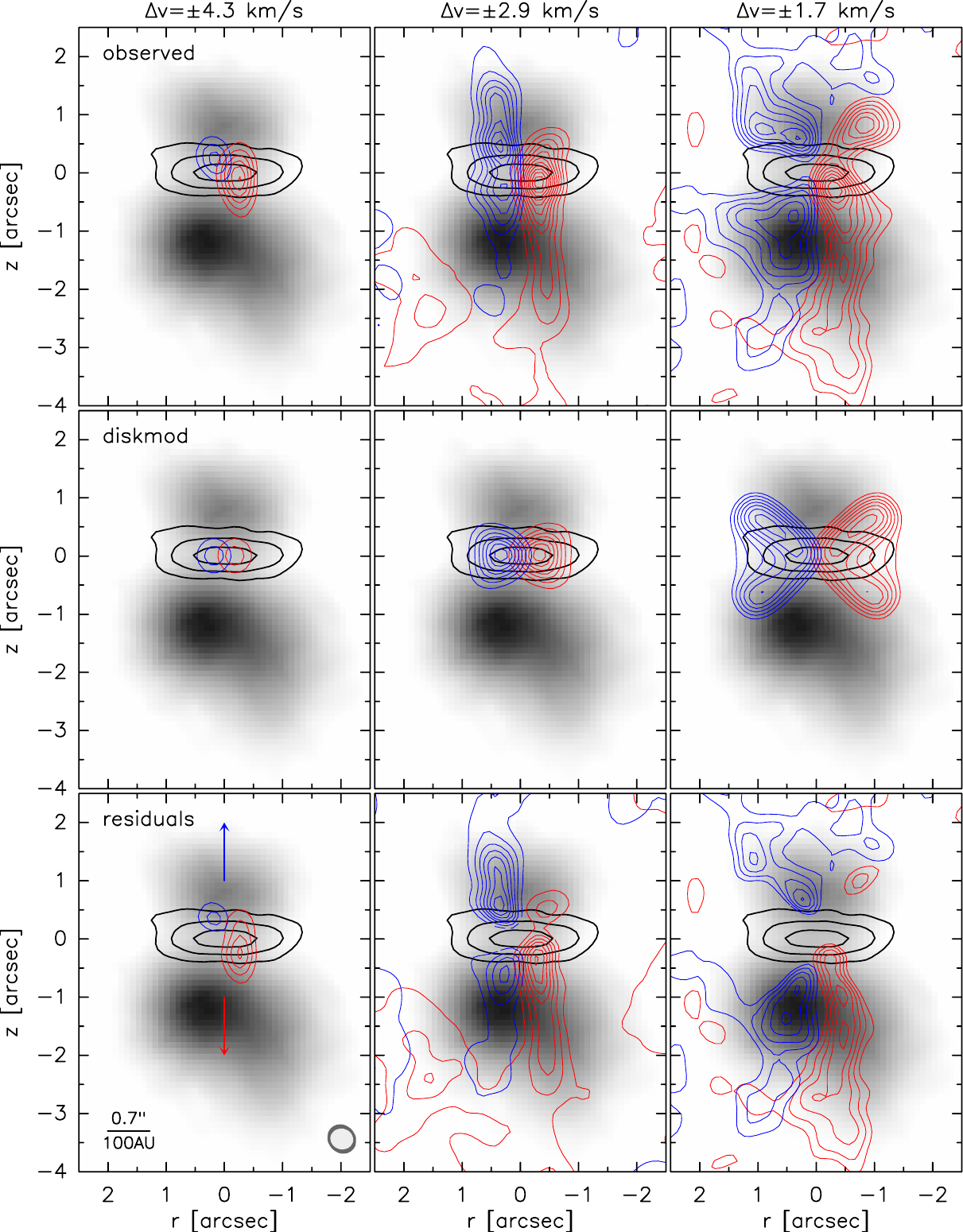}
\caption{\label{fig_binchan}
Channel maps of the $^{12}$CO\,(2--1) emission in binned, 0.8\,\kms\ wide channels around the velocity offsets $\Delta v$\ from $v_0=5.95$\,\kms\ indicated on top.
Top row: observed total emission (see Fig.\,\ref{fig_chanmap_12co_obs}); 
middle row: disk model (see Sect.\,\ref{ssec:mod:disk} and Fig.\,\ref{fig_chanmap_12co_mod});
bottom row: residuals, that is, observed total emission with disk model subtracted.
Blue contours indicate negative velocity offsets, red contours positive ones. CO contours start at 15\,mJy/beam (left and middle panels), and 40\,mJy/beam (right panels) and are spaced by 15\,mJy/beam. Blue and red arrows in the lower left panel show the inclination orientation of the disk and outflow. Black contours show the 230\,GHz dust continuum emission from the disk at 2.5, 7.5, and 15\,mJy/beam. Synthesized FWHM beam sizes are shown as gray ellipses in the lower right corner of the lower left panel (dark gray: CO, light gray: dust). 
The underlying gray-scale image shows the NIR K-band image of the bipolar reflection nebula (see Fig.\,\ref{fig-overview}).
 }
\end{center}
\end{figure}

\begin{figure}
\begin{center}
\includegraphics[width=0.48\textwidth]{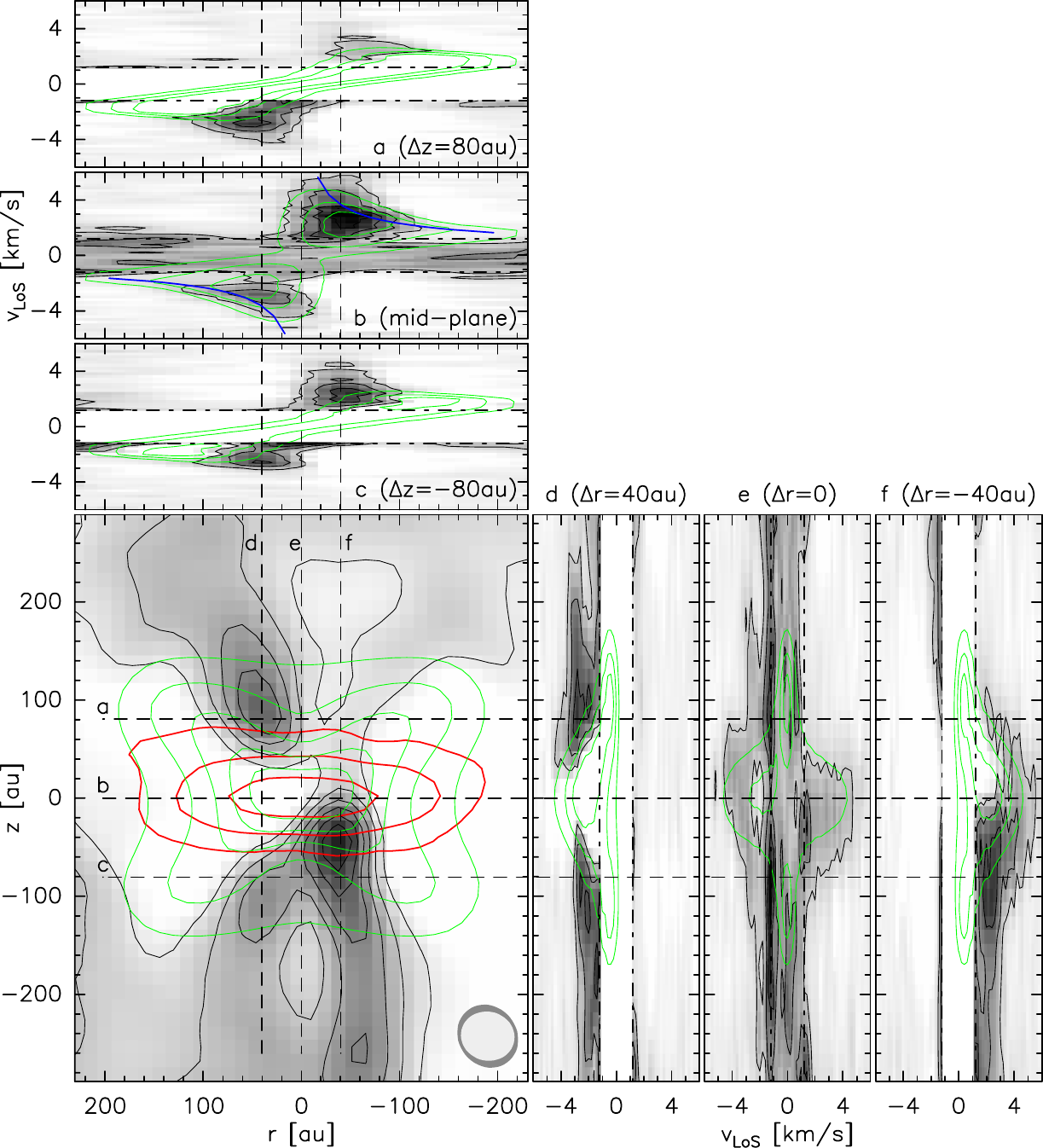}
\caption{\label{fig_pvddiff}
Bottom left: total intensity map (gray-scale and black solid contours) of the residual $^{12}$CO\,(2--1) emission after subtracting the disk model and masking the corrupted central velocity channels between $\pm1.2$\,\kms, integrated over the velocity range $\pm6$\,\kms. Red contours show the observed 230\,GHz dust continuum emission from the disk at 2.5, 7.5, and 15\,mJy/beam. Green contours show the total integrated intensity of the CO emission from the model disk. Synthesized FWHM beam sizes are shown as gray ellipses in the lower right corner (dark gray: CO, light gray: dust).~~
Top: position-velocity diagrams (PVD) of $^{12}$CO\,(2--1) at three cuts parallel to the plane of the disk (a, b, c, as indicated by dashed lines in the total intensity map; gray-scale and black contours). The middle panel shows the PVD of the unmasked total observed CO emission along the disk mid-plane, while the upper and lower panels show disk wind PVDs of the residual CO emission (after subtracting the disk model) and with the central corrupted velocity channels masked. Green contours show the PVDs of the disk CO model along the disk mid-plane.~~
Bottom right: same as above, but for the three vertical cuts labeled d, e, and f. Green contours show the PVDs of the disk CO model along the same vertical cuts.
}
\end{center}
\end{figure}

\begin{figure*}
\begin{center}
\includegraphics[width=1.0\textwidth]{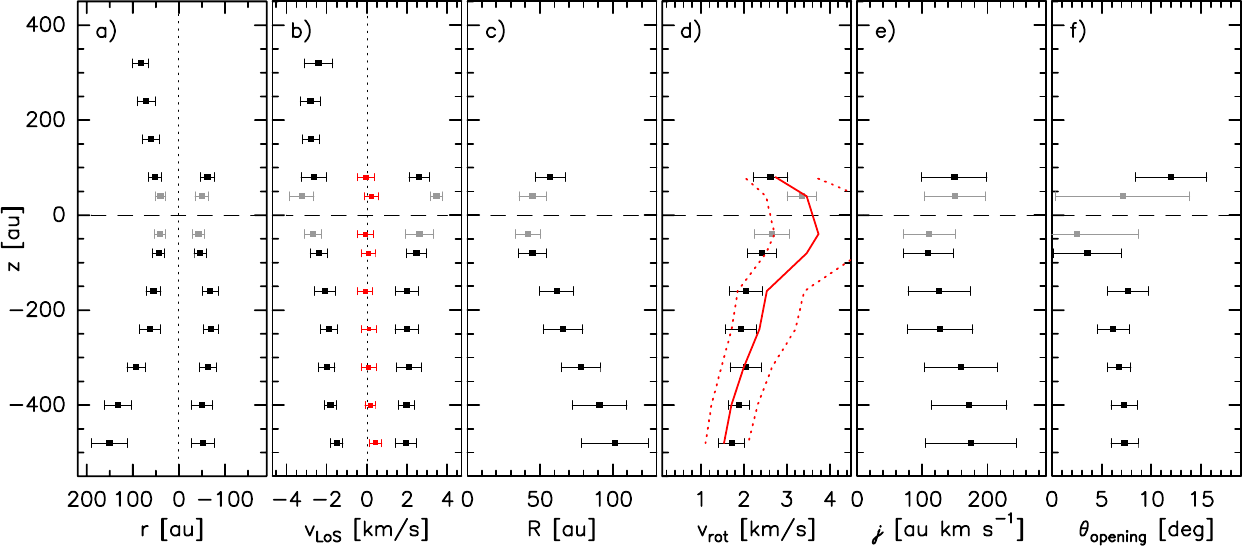}
\caption{\label{fig_pvddiff_means}
Outflow parameters derived from PVDs of disk-subtracted CO residuals at different vertical heights below and above the disk midplane:\\
{\it a)} X location $r$\ and FWHM width of the outflow cone walls. The errorbars denote the FWHM widths of the cone walls, not the uncertainty of the mean (which is smaller than the symbols). The light gray data points at $z=\pm40$\,au might be affected by imperfections in the disk model that was subtracted from the data.~~
{\it b)} Line-of-sight (LoS) velocity (relative to systemic velocity $v_0$) and FWHM width of the outflow cone walls. Red squares show the center velocity between left and right outflow cone.~~
{\it c)} Radius $R$\ of the outflow cone, with uncertainties derived from the FWHM widths of the outflow cone walls.~~
{\it d)} Rotation velocity and 1\,$\sigma$\ uncertainty of the outflow cone walls. The red solid and dotted lines show the expected rotation velocity and uncertainty for angular momentum conservation of a wind that is launched from the disk at a single radius $R_\mathrm{conn}=35\pm10$\,au with a Keplerian rotation velocity of $v_\mathrm{K}=3.7\pm0.3$\,\kms.~~
{\it e)} Specific angular momentum $j$\ and 1\,$\sigma$\ uncertainty.~~
{\it f)} Half-opening angle $\theta$\ and 1\,$\sigma$\ uncertainty.
}
\end{center}
\end{figure*}

%
Here we analyze the channel maps and PVDs of the residual \mbox{$^{12}$CO(2--1)} emission, that is, after subtracting the CO disk model, and derive the outflow parameters. As for the disk model, we use the rotated (by 32\degr, Sect.\,\ref{ssec:res:ov}) maps such that the disk midplane is aligned with the x-axis, and the outflow and disk rotation axis is aligned with the z-axis. Following \citet[][their Fig.\,4]{hirota2017}, we use cylindrical coordinates and denote the axis parallel to the disk midplane $"r"$, and the perpendicular rotation axis $"z"$.
The high-resolution CO data used here to analyze the detailed structure of the outflow as close as possible to its origin reliably trace the outflow cone walls only out to $\approx$600\,au below the disk. But, the integrated intensity maps, especially at lower angular resolution (Fig.\,\ref{fig_covelfield}), indicate a total length of the south eastern outflow lobe of $\approx$1100\,au, which we adopt here as the total half-length of the CO outflow.

Figure\,\ref{fig_velfield} shows the mean velocity fields ($1^{\mathrm st}$\ moment maps) of $^{12}$CO of the best-fit disk model and of the residual (disk-subtracted) CO emission, overlaid with contours of the integrated line emission ($0^{\mathrm th}$\ moment). The butterfly-shaped appearance of both the total intensity map and the mean velocity field already indicate that the residual emission resembles a weakly opened double cone that connects to the disk at $r\approx 40$\,au and rotates with the same orientation as the disk. 

This is even more clearly seen in the binned channel maps (Fig.\,\ref{fig_binchan}). At the highest rotation velocities with \mbox{$\Delta v=\pm4.3$\,\kms}, we see emission from the innermost parts of the Keplerian disk (middle left panel), but also already disk wind residuals emerging nearly perpendicular off the disk surface at \mbox{$\Delta r\approx \pm0\dotsec2-0\dotsec25$}\ (30--40\,au). At intermediate rotation velocities with $\Delta v=\pm2.9$\,\kms, the disk wind residuals in the lower middle panel show long linear streaks extending nearly perpendicular from the disk with a very small opening angle in both directions and both up and downward at $\Delta r\approx\pm40$\,au.
At the lowest rotation velocities with $\Delta v=\pm1.7$\,\kms, the disk wind residuals seem to suggest a slightly larger opening angle, in particular at the upward-facing blue-shifted left wing. But this impression may, at least in part, be caused by the not perfect separation of emission from the disk atmosphere from the disk wind emission. The lower side of the outflow appears to be slightly bent toward the left at heights $z<-200$\,au below the disk.
The upper red-shifted (right) cone wall is in all three velocity ranges the least prominent wing as it may be most affected by the combination of red-shift due to rotation and blue-shifted outflow direction (facing toward us), and thus be blended by the self-absorbing envelope. The same, but to a lesser degree, applies to the lower blue-shifted (left) cone wall.

Figure\,\ref{fig_pvddiff} shows the total intensity map of the residual \mbox{$^{12}$CO\,(2--1)} emission after subtracting the disk model and masking the corrupted central velocity channels between $\pm1.2$\,\kms, together with the total intensity contours of the CO emission from the disk atmosphere and the thermal dust emission from the disk. The X-shaped morphology of the outflow with pronounced upper left (blue-shifted) and lower-right (red-shifted) wings becomes again very evident. The PVD of the horizontal cut through the disk mid-plane (top) shows the excellent, albeit not perfect match between the total observed CO emission and the CO model of the disk. The PVDs of the horizontal cuts through the residual (disk-subtracted) CO emission at $z=\pm80$\,au show that there is clearly excess emission at $\Delta r\approx\pm40$\,au from the disk rotation axis and at $\Delta v_{\rm LoS}\approx\pm3$\,\kms. The PVDs of the vertical cuts through the residual CO emission at $\Delta r=\pm40$\,au show that the LoS velocity of the CO emission from the vertical streaks does not significantly change (decrease) with separation from the disk.


To derive more quantitative outflow parameters, such as the outflow width, opening angle, and specific angular momentum distribution, we generate PVDs, such as in Fig.\,\ref{fig_pvddiff}, at different heights $z=-480$\ to 320\,au by 80\,au  below and above the disk and apply Gaussian fits to cuts through the wind cone emission along the $X$\ axis, and visual inspection of cuts along the velocity axis (to avoid confusion with the residual disk and envelope emission). 
This approach corresponds to the "double-peak separation method" described by \citet{tabone2020}, with the only difference that we determine the offsets and FWHM by fitting 1-D profile cuts through the PVD, instead of just locating the peak in the PVD image.
At vertical heights \mbox{$| z|\geq320$\,au}, the cone wall profiles are no longer approximately Gaussian and symmetric, but show a broad wing of emission toward the inner side of the cone. Here, we only fit the core of the emission profile. Figure\,\ref{fig_pvddiff_means} shows the resulting values for the X-location, $r$, and mean LoS velocity, $v_{\rm LoS}$, of the outflow cone walls, as well as the radius, $R$, the toroidal or rotation velocity, $v_{\rm rot}$, specific angular momentum, $j$, and half-opening angle, $\theta$, as function of vertical distance from the disk midplane, $\Delta z$.

The right panel {\it a)} of Fig.\,\ref{fig_pvddiff_means} shows that the outflow remains strongly collimated out to $z\approx-500$\,au below the disk. Above the disk, the emission from the right cone wall is only traced reliably out to $z\approx100$\,au, possibly due to the reasons mentioned above. Therefore, it is unfortunately impossible to derive the outflow width, rotation velocity, etc., at larger $z$\ above the disk. However, the location and LoS velocity of the upper left cone wall suggests that the upper (blue-shifted) lobe behaves very similar as the lower (red-shifted) lobe, at least out to distances of $z\approx-500$\,au, until which the left cone wall is traced. Furthermore, it is evident that the entire outflow seems slightly bent toward the left at vertical distances $z\lesssim-200$\,au below the disk. We take a very conservative approach to estimating the uncertainties from the FWHM widths of the cone walls and their velocity, rather than using the formal uncertainties of the Gaussian profile fits, which would result in unrealistically small error bars. Although this may lead to an overestimation of the actual uncertainties, this approach should include the errors introduced by the self-absorption by the envelope, as well as the imperfect disk model subtraction from the data. We note that the cone wall FWHM widths are all smaller than the synthesized beam size of $\approx$70\,au, which implies that the cone walls are spatially unresolved.

Panel {\it b)} shows the LoS velocity of the left (blue-shifted, that is, negative velocities) and right (red-shifted) cone walls. The two velocities stay approximately symmetric with respect to the systemic velocity of the disk, while the absolute LoS velocity offsets slowly decrease with vertical distance from the disk. What is remarkable, though, is that there is no significant jump in the mean velocity between the upward, on large scales blue-shifted, and downward, on large scales red-shifted, outflow wings. Only at distances $z\le-320$\,au downward, there is a slight shift of -0.8\,\kms. If we use the mean velocity offset between the upward and downward-facing cone walls of $\pm$0.12\,\kms, we derive with $i=87\pm2$\degr\ a radial outflow velocity of $v_0=2.3^{+6.9}_{-1.4}$\,\kms. However, this analysis is hampered by the fact that the emission from the right upward-facing (blue-shifted) cone wall is only traced reliably out to $z\approx100$\,au. 
We therefore consider this value a lower limit to the radial outflow velocity $v_0$.
On the other hand, we derive in \citet{launhardt09} from the lower-resolution CO maps a mean LoS velocity offset of $\pm$0.65\,\kms\ by cross-correlating the mirrored PVDs of the two outflow lobes. Using $i_{\rm disk}=87\pm2$\degr, we derive $v_0\approx12^{+25}_{-6}$\,\kms, which we consider an upper limit. In lack of more precise constraints, we therefore adopt a mean value of $v_0=7^{+20}_{-5}$\,\kms\ for the radial outflow velocity.

Panel {\it c)} shows the radius of the outflow cone \mbox{$R=r^{\rm left}-r^{\rm right}$}, which increases slowly with separation from the disk from \mbox{$R\approx 43\pm5$\,au} at $z =\pm40$\,au to $R\approx$\,102\,au at $z=-480$\,au. The upward-facing cone is only fully traced out to $z=80$\,au, where we determine a cone radius of $R\approx 57$\,au. Seeing that the cone radius increases approximately linearly with $z$, we extrapolate a launch radius of the cone walls at $z\approx40$\,au, where the surface of the flared disk atmosphere at this radius is (see Fig.\,\ref{fig-mod-disk}), of \mbox{$R_{\rm L}=40\pm5$\au}.
The observed thickness of these wind cone walls is spatially unresolved (i.e., $\ll$70\,au), and there is no direct indication in the channel maps that the wind is launched from a wider range of radii. However, given the angular resolution of our data, we cannot exclude that the main launching region extends over a few tens of au on the disk surface, that is, basically from the inner rim of the CO disk at $R_{\rm in}^{\rm gas}\approx12$\,au to the CO iceline at $\approx$80\,au (Sect.\,\ref{ssec:mod:disk}).

Panel {\it d)} shows the rotation velocity\footnote{The factor $1/v\sin(i)$\ can be neglected here since $\sin(87\degr)\approx 0.999$.}, \mbox{$v_{\rm rot}=v_{\rm LoS}^{\rm right}-v_{\rm LoS}^{\rm left}$}, of the outflow cone walls, which gradually decreases from \mbox{$v_{\rm rot}\approx 3.0\pm0.35$\,\kms} at \mbox{$z=\pm40$\,au to $\approx$1.7\,\kms} at \mbox{$z=-480$\,au}. Considering a central mass of 0.55$\pm$0.1\,\msol\ and the increasing radius $R$\ of the outflow cone with vertical distance $z$\ from the disk, this decreasing rotation velocity is consistent with (specific) angular momentum conservation \mbox{$j=R\times v_{\rm rot}\approx140\pm33$\,au\,\kms} of a flow that is launched from the disk surface at the single radius $R_{\rm L}=40$\,au with the Keplerian rotation velocity of \mbox{$v_{\rm K}\approx 3.5\pm0.4$\,\kms}.  

Panel {\it e)} shows the specific angular momentum, \mbox{$j=R\times v_{\rm rot}$}, and its 1\,$\sigma$\ uncertainty. Although it seems that $j$\ is slowly increasing with $z$\ from 110\,au\,\kms\ at $z=-40$\,au to 175\,au\,\kms\ at -480\,au, this is significant only at the 2-$\sigma$\ level.

Panel {\it f)} shows the half-opening angle, $\theta = \arctan \left(\frac{R-R_{\rm L}}{z}\right)$. Below the disk, where we can trace the cone walls out to $z \approx 500$\,au, the opening angle appears to be approximately constant with $\theta \approx 7\degr$. Above the disk, where we can trace the cone walls out to only $z \approx 160$\,au, the opening angle appears to be slightly larger with $\theta\approx14\pm2\degr$.

\begin{table*}[ht]
\caption{Derived parameters of the CO outflow}
\label{tab-OFpar} 
{\footnotesize  
\begin{tabular}{lllll}
\hline \hline
Parameter & Symbol & Value & Unit & Remark/reference  \\
\hline 
Outflow half length           & $r_0$     & $1100\pm200$ & au  & Sect.\,\ref{ssec:mod:outflow1} \\[1mm]
Launch radius           & $R_{\rm L}$ & 20\,--\,45 & au & Sects.\,\ref{ssec:mod:outflow1} and \ref{ssec:dis:outflow}\\[1mm]
Half-opening angle            & $\alpha$  &  $7\pm2$  & deg & Fig.\,\ref{fig_pvddiff_means}  \\[1mm]
Radial outflow velocity & $V_0$     & $7^{+20}_{-5}$  & \kms & Sect.\,\ref{ssec:mod:outflow1} \\[1mm]
Dynamical outflow time  & $\tau_{\rm dyn}$ & $740_{-500}^{+2000}$ & yr & $r_0/V_0$ \\[1mm]
Total outflow mass            & $M_{\rm CO}$  & $(1.0\pm0.3)\times 10^{-3}$ & \msol & Sect.\,\ref{ssec:mod:outflow1}\tablefootmark{a} \\[1mm]
Mass flux               & $\dot{M}_{\rm CO}$ & $\left(1.4^{+3.7}_{-1.0}\right)\times10^{-6}$ & \msol\,yr$^{-1}$  & $M_{\rm CO}/\tau_{\rm dyn}$  \\[1mm]
Total outflow momentum  & $P_{\rm CO}$  & $\left(7_{-5}^{+20}\right)\times10^{-3}$ & \msol\,\kms & $M_{\rm CO}\times V_0$ \\[1mm]
Momentum flux (thrust)  & $\dot{P}_{\rm CO}$ & $\left(1.0_{-0.9}^{+2.6}\right)\times10^{-5}$  & \msol\,\kms\,yr$^{-1}$  & $P_{\rm CO}/\tau_{\rm dyn}$ \\[1mm]
Total angular momentum  & $J_{\rm CO}$  & $(2.1\pm1.0)\times10^{7}$  & \msol\,km$^2$\,s$^{-1}$  & $M_{\rm CO}\times j_{\rm CO}$  \\[1mm]
Angular momentum flux & $\dot{J}_{\rm CO}$  & $\left(2.8_{-2.3}^{+7.5}\right)\times10^{4}$ & \msol\,km$^2$\,s$^{-1}$\,yr$^{-1}$  & $J_{\rm CO}/\tau_{\rm dyn}$  \\[1mm]
Specific angular momentum & $j_{\rm CO}$ & $140\pm40$ & au\,\kms & $R\times V_{\rm rot}$, Fig.\,\ref{fig_pvddiff_means} \\
\hline
\end{tabular} \tablefoot{
\tablefoottext{a}{$\Sigma^{\rm gas}=\Sigma_0^{\rm gas}\,\left(\frac{r}{1\,{\rm AU}}\right)^p$}
}}
\end{table*}

\begin{figure}
\begin{center}
\includegraphics[width=0.49\textwidth]{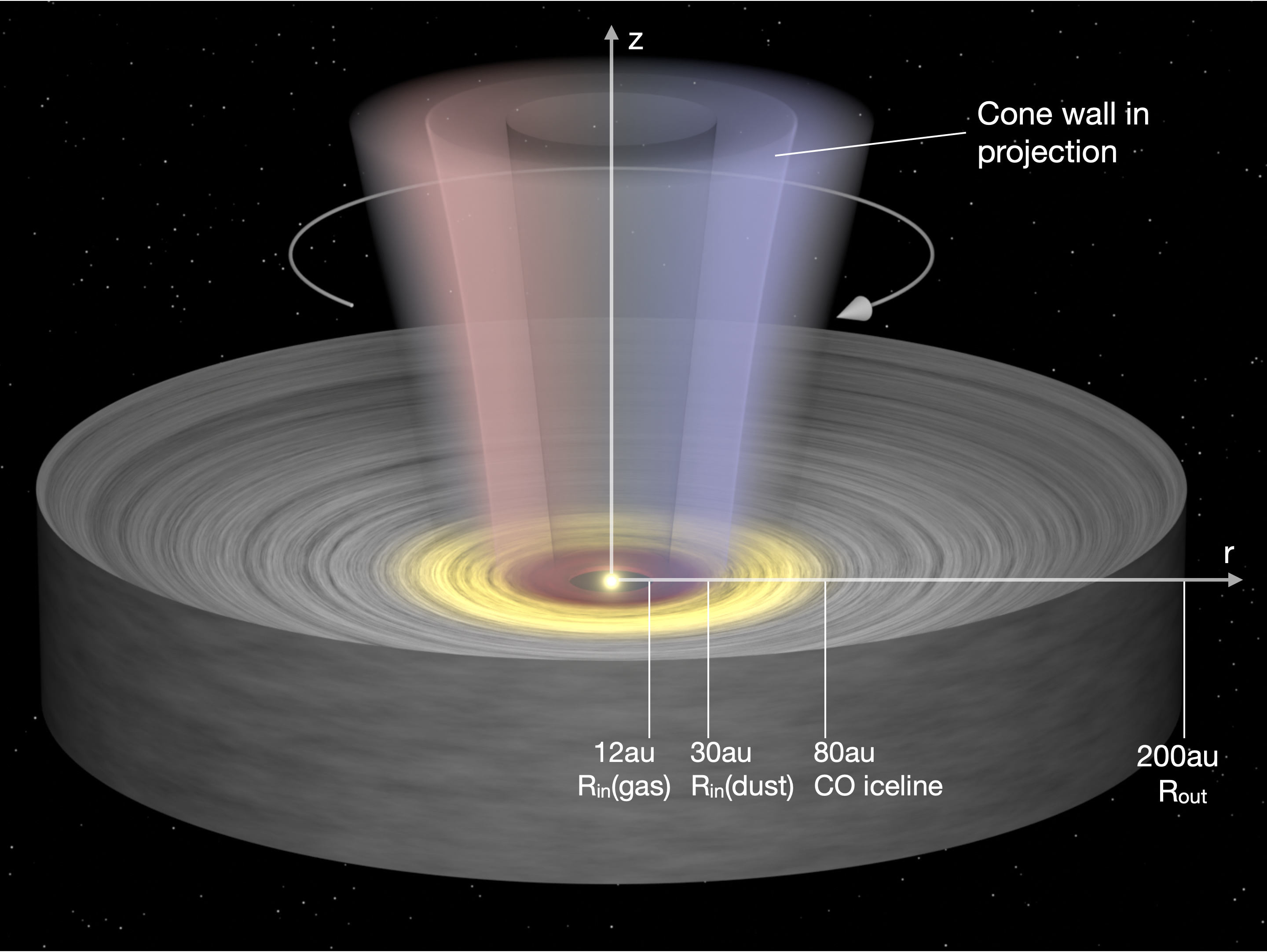}
\caption{\label{fig_sketch}
Purely empirical schematic cartoon illustrating the main features and geomoetry of the disk and disk wind derived in this paper.
}
\end{center}
\end{figure}

%
As already mentioned in Sect.\,\ref{ssec:res:12co}, the new high-resolution CO map recovers the same total flux as seen in the lower-resolution map of \citet[][$\approx$35\,Jy\,\kms]{launhardt09}. Of this, 10\% (3.5\,Jy\,\kms) originate from the disk atmosphere, $\approx$19\% (6.5\,Jy\,\kms) from the disk wind cone walls, and 71\% (25\,Jy\,\kms) from the diffuse component. To be more precise, what we call here the "wind cone walls" are only the edge-on-seen parts of the cone walls, while the "diffuse component" is mostly related to the other, in projection less-structured parts of the cone walls, such that the total flux from the disk wind amounts to $31.5\pm3$\,Jy\,kms. However, due to the masked central velocity channels at $\Delta v=\pm1.2$\,\kms\ (Sect.\,\ref{ssec:res:12co}), this value may slightly underestimate the actual total CO flux.

Making the simplifying assumption that most of the CO emission from the outflow is optically thin (which may not be strictly fulfilled for the cone walls), assuming local thermodynamic equilibrium, and following the formalism outlined in \citet[][their eq.\,A7]{bourke1997}, here rewritten for a standard value of $X=N({\rm H}_2)/I({\rm CO})=2.3\times10^{20}$\,cm$^{-2}$/(K\,\kms) \citep[e.g.][]{rohlfs1996}, standard ISM composition with mean molecular weight $\mu_{\rm m}=1.36$, and substituting $T_{\rm mb}=\frac{\lambda^2\,F_{\nu}}{2\,k\,\Omega}$, the total gas mass, often referred as the CO mass, is given by 
\begin{equation}
\left(\frac{M_{\rm CO}}{\rm M_{\odot}}\right) \approx 8\times10^{-4}\,\left(\frac{D}{\rm kpc}\right)^2\,\left(\frac{\lambda}{\rm mm}\right)^2\,\left(\frac{\int I_{\nu}d\nu}{\rm Jy\,\kms}\right) ~~,
\label{eq:mco}
\end{equation}
where $\Omega$\ is the source solid angle, $D$\ is the source distance in kpc, and $\lambda$\ is the observing wavelength. Using $\Omega\approx40$\,arcsec${^2}$, \mbox{$D=0.14$\,kpc}, $\lambda=1.3$\,mm, and \mbox{$\int I_{\nu}d\nu=31.5$\,Jy\,\kms}, we derive a total mass of the visible outflow of $8.4\times10^{-4}$\,\msol, which we round up to \mbox{$M_{\rm CO}\approx(1\pm0.3)\times10^{-3}$\,\msol} because of the CO flux under-estimation mentioned above. This is basically the same value as was derived by \citet{launhardt09} by integrating the CO model of the outflow. Table\,\ref{tab-OFpar} summarizes this and other derived parameters of the CB\,26 CO outflow. The schematic cartoon shown in Fig.\,\ref{fig_sketch} summarizes the geometry of the disk and disk wind derived here.


\section{Discussion} \label{sec:dis}


\subsection{The disk and central star} \label{ssec:dis:disk}

According to its spectral index $\alpha_{\nu}^{(2.2-25\,\mu{\rm m})}\approx-3.8$, CB\,26-YSO\,1 can be classified as a Class\,I embedded YSO \citep{adams1987}.
Thanks to the relatively tight constraint on the mass of the central star, evolutionary track models provide reasonable constraints on the age of the system, despite the relatively large uncertainty of its luminosity. While a main-sequence star of $M_{\ast}=0.55\pm0.1$\,\msol\ (M0.5\,V) would have a bolometric luminosity of $L_{\ast}\approx0.05$\,\lsol, a star with this mass with $L_{\ast}=1.0\pm0.4$\,\lsol\ (App.\,\ref{sec:app:lstar}) should have an age of $1\pm0.5$\,Myr, depending on the model used \citep[e.g.,][]{Siess_etal2000,allard2014,baraffe2015}. The effective temperature of such a star should be $T_{\rm eff}=3700\pm150$\,K. Even if we consider the unlikely case of the mass and luminosity split into two equal-mass stars, the mean (most likely) age would still be $\approx$1.5\,Myr. 
On the other hand, \citet{simon2019} have shown that, when using using magnetic stellar evolution isochrones from, for example, \citet{feiden2016}, the inferred ages of young stars (1-10\,Myr) can be up to three times larger than the ones derived from nonmagnetic models. Considering this effect, we adopt an age of $1^{+2}_{-0.5}$\,Myr for CB\,26-YSO1.
We also tried to derive further constraints on the spectral type(s) of the central star(s) by employing visual long-slit spectroscopy on the bipolar reflection nebula, hoping that we could recover at least the strongest stellar lines, but this was not successful.

The size of central hole in the dust disk, which was independently derived by several authors (see Table\,\ref{tab:diskpar}), has a relatively large uncertainty with values (incl. uncertainties) ranging from $R_{\rm in}^{\rm dust}\approx$10\,au to $\approx$50\,au. The value and uncertainty derived in the most recent and comprehensive study and based on the most extensive data set, $R_{\rm in}^{\rm dust}=16^{+37}_{-8}$\,au \citep{zhang2021} actually overlaps with $R_{\rm in}^{\rm gas}\approx12\pm6$\,au, which we derive from the CO emission (Sect.\,\ref{ssec:mod:disk}, Table\,\ref{tab-diskpar}). 

An inner hole in the detected dust emission does not necessarily mean that there is a sharp physical cut-off at this radius, but it could also result from a combination of decreased dust opacity and column density. For the gas emission, the steep Keplerian velocity gradient at small radii spreads out the emission over a steeply increasing velocity range. Furthermore, spatial "beam dilution" can lead to a significant loss of sensitivity to emission from the inner disk, where the ratio of emitting area to beam size becomes increasingly smaller, in particular for HCO$^+$\ with the relatively large synthesized beams of $\approx$3\asp5 for HCO$^+$(1--0) (corresponding to 490\,au linear resolution) and $\approx$1\asp35 for HCO$^+$(3--2) (190\,au). Hence, the lack of detected emission from Keplerian velocities corresponding to radii $<$80\,au (Sect.\,\ref{ssec:res:hco}) does not necessarily mean that there is a HCO$^+$\ hole of that size in the disk. The same argument also holds for the CO emission, that is, the value of $R_{\rm in}^{\rm CO}\approx12\pm6$\,au, derived in Sect.\,\ref{ssec:mod:disk}, may as well be affected by this spectral and spatial beam dilution and the CO disk might have a much smaller or even no inner hole at all. Molecular line observations of CO and HCO$^+$\ at higher angular resolution and better sensitivity would be required to better constrain the existence and size of the inner whole in the gas disk.
Despite these uncertainties, the CB\,26 disk with such a significant inner hole in the dust emission can be considered as a transitional disk \citep{strom1989,skrutskie1990,espaillat2014}, although the primary cause for this inner disk clearing remains undisclosed. We don't know whether it is caused by a stellar companion, newly formed giant planets, or photoevaporation \citep[e.g.][]{alexander2014,espaillat2014,pascucci2023}.

The mass of the flared disk is also uncertain with estimates ranging from $M_{\rm disk}\approx 0.08$\,\msol\ \citep{zhang2021} to $\approx$0.2\,\msol\ (this paper, Sect.\,\ref{ssec:mod:disk}), and $\approx$0.3\,\msol\ \citep{sauter09}. These differences can, at least in part, be attributed to the use of different dust opacities. Considering that the latter estimate seems unrealistically high, we therefore adopt a value of \mbox{$M_{\rm disk}\approx0.15\pm0.05$\,\msol}. With this, the disk mass amounts to $\approx27\pm10$\% the mass of the central star. The disk would thus be prone to gravitational instabilities and developing quasi-stable spiral arms that process infall from the surrounding cloud \citep[e.g.,][see also Sect.\,\ref{ssec:dis:env}]{kratter2016}. 


\subsection{The envelope} \label{ssec:dis:env}

We actually do not see any emission from the extended envelope in our interferometric molecular line and thermal dust continuum maps used in this paper. The reason could be that such extended emission is simply resolved out in the interferometric maps. But we clearly see the effects of self-absorption due to cool CO (and HCO$^+$) gas in the envelope. While the fast-rotating inner parts of the Keplerian disk are not affected, the CO emission from the slower-rotating outer parts might still be affected by this self-absorption. The size and total mass of the envelope was estimated from single-dish observations to be $r_{\rm env}\approx10^4$\,au and \mbox{$M_{\rm env}\approx0.1-0.2$\,\msol} \citep{launhardt10,launhardt13}. The total gas mass of the globule CB\,26, in which \mbox{CB\,26\,-\,YSO\,1} is embedded, was estimated to be $\approx$2.4\,\msol\ \citep{launhardt13}. 

Since the CB\,26 YSO and disk are still embedded in a denser envelope, which itself is embedded in the globule CB\,26 (Fig.\,\ref{fig-overview}), it is very likely that the disk still accretes gas from the envelope, even though we have no direct observational evidence of such late accretion. As \citet{kuffmeier2023} have shown, such late accretion can occur well beyond an age of 1\,Myr, and it can still transport significant amounts of mass and angular momentum onto the disk. This could explain the relatively large disk mass ($\approx27\pm10$\% the mass of the central star; Sect.\,\ref{ssec:dis:disk}), which is expected to reach its maximum during this phase \citep{hogerheijde2001}. The disk wind then just carries away this excess angular momentum.


\subsection{The outflow} \label{ssec:dis:outflow}

The disk wind is launched from the surface of the flared disk with a nearly constant half-opening angle $\theta\approx$\,7\degr, leading to a rotating cone that continuously opens from radius $\approx$\,45\,au at $|\Delta z|=80$\,au to $\approx$\,90\,au at $|\Delta z|=400$\,au and can be traced by our high-resolution CO maps until about 300\,au above, and 600\,au below the disk midplane. The reason that the cone walls cannot be traced further away from the star and disk by the high-resolution data analyzed here is probably related to a combination of self-absorption by the envelope, decreasing contrast between the widening cone walls and ambient material, as well as sensitivity.

The analysis of the residual CO emission (Sect.\,\ref{ssec:mod:outflow1}), after subtracting the modeled emission from the disk atmosphere, suggests a launch radius at the disk surface of \mbox{$R_{\rm L}=40\pm5$\au}, which corresponds to the inner part of the optically thick disk ($R_{\rm in}^{\rm dust}\approx 16^{37}_{-8}$\,au, $R_{\rm out}^{\rm dust}\approx200\pm50$\,au). 
Since this value was derived by linear extrapolation of the cone wall locations from $|z|>80$\,au inward and with no further observational constraints of possible collimation acting inside $|z|<80$\,au, we consider it to represent an upper limit to the mean launch radius.
We compare this with the ejection radius predicted by \citet[][their eq.\,4]{Anderson03} for the launching radius of a cold, steady, axisymmetric MHD disk wind at large distances, where the gravitational potential of the star is negligible, in the form derived by \citet{Pety_etal2006}:
\begin{equation}
R_{\rm ML} = R\,\left(2\,\frac{V_{\rm Kep}\,V_{\rm rot}}{V_0^2} \right)^{2/3}~,
\label{eq:rml}
\end{equation}
where $R$\ is the distance from the outflow axis to the considered volume element, $V_{\rm Kep} = \sqrt{G\,M_{\ast}/R}$\ is the Keplerian velocity at this position, and $V_{\rm rot}$\ and $V_0$\ are the measured rotation and radial outflow velocities at that position. 
Using the values and uncertainties for $R$\ and $V_{\rm rot}$\ from Fig.\,\ref{fig_pvddiff_means}\ for \mbox{$z=-240\ldots-480$\,au}, and a constant (with $z$) radial outflow velocity $V_0=7^{+20}_{-5}$\,\kms, we derive consistently \mbox{$R_{\rm ML}\approx28^{+15}_{-10}$\,au}. This value is only slightly smaller than, and overlaps in the error margin with the launch radius $R_{\rm L}$\ derived directly from the data, and we consider it as a lower limit to the mean launch radius. Therefore, we finally adopt $R_{\rm L}\approx20-45$\,au as the most likely range for the mean wind launch radius.

Despite its large uncertainty, the radial lift-off wind velocity of $7_{-5}^{+20}$\,\kms\ (Sect.\,\ref{ssec:mod:outflow1}) is still is significantly smaller than the tangential velocity of HH\,494 (\mbox{$86\pm17$\,\kms}, App.\,\ref{sec:app:hh494}), suggesting that an as-yet undetected high-velocity jet, rather than the outer disk wind, is responsible for HH\,494. In App.\,\ref{sec:app:hh494}, we derive a dynamical age for this HH object-exciting jet of $\approx$3000\,yrs, which could imply that a, possibly FU Orionis-like \citep{hartmann1996} accretion event a few thousand years ago must have filled the inner disk with gas (if there was a gas hole already present at this time) and triggered an energetic inner disk wind that produced this jet. Unfortunately, we could not derive any constraints on the current mass accretion rate onto the star, if there still is any, due to the extreme edge-on morphology.

With the launch radius and the specific angular momentum of the disk wind at hand, we can derive the magnetic lever arm parameter, $\lambda$:
\begin{equation}
\label{eq:lambda}
\lambda = \frac{j}{\Omega_0\,R_{\rm L}^2}~,
\end{equation}
where $j$\ is the mean specific angular momentum of the outflowing gas, and $\Omega_0$\ is the Keplerian angular rotation speed at the launch point $R_{\rm L}$\ \citep{blandford1982,spruit1996,ferreira2006,tabone2020}. Using the values and uncertainty ranges listed in Table\,\ref{tab-OFpar}, we derive $\lambda\approx 1.3\pm0.7$. Such a small value of $\lambda$, which is an essential feature of MHD wind models that relates the mass outflow rate to the accretion onto the disk, would imply a very efficient extraction of angular momentum, but is in line with recent nonideal MHD simulations of magnetized winds from protoplanetary disks \citep[see][and references therein]{devalon2020}, as well as with observational constraints derived for other rotating disk winds in recent years (see Sect.\,\ref{ssec:dis:comp} and Table\,\ref{tab-ofcomp}). Vice versa, this efficient extraction of angular momentum by the disk wind implies that the angular momentum flux by accretion onto the disk, for which we have no direct observational constraints, must be of the same order as the one carried away by the disk wind. that is, a few times $10^4$\,\msol\,km$^2$\,s$^{-1}$\,yr$^{-1}$\ (Table\,\ref{tab-OFpar}).
  
An important parameter for gaining insights into the wind-launching mechanism is the observed ratio of the outflow momentum flux (thrust) to the maximal possible thrust that can theoretically be provided by the bolometric luminosity of the central star \citep{pudritz2007}. In Sect.\,\ref{ssec:mod:outflow1} (Table\,\ref{tab-OFpar}), we derive a total outflow momentum flux of  \mbox{$\dot{P}_{\rm CO} = (1.0_{-0.9}^{+2.6})\times10^{-5}$\,\msol\,\kms\,yr$^{-1}$}. The maximum thrust that can be provided by a central star with bolometric luminosity $L_{\ast}=(1\pm0.4)$\,\lsol\ is \mbox{$L_{\ast}/c= (2\pm0.8)\times10^{-8}$\,\msol\,\kms\,yr$^{-1}$}, which is nearly three orders of magnitude smaller than the CO outflow thrust. Hence we conclude that photoevaporation cannot be the main driving mechanism for this outflow, but it must be predominantly a MHD-driven disk wind, even if we have no observational constraints on the morphology and strength of the local magnetic field\footnote{From polarimetric submillimeter dust continuum observations, \citet{henning01} derived a mean magnetic field strength of \mbox{$B_{\rm CB26}\approx74\,\mu$G} in the envelope, oriented along \mbox{P.A.$\approx$25\degr}, but they could not spatially resolve the local structure of the magnetic field around the disk.}. Indeed, the CB\,26 disk wind falls right on the relation $F_{\rm CO}/F_{\rm rad}=250\,(L_{\rm bol}/10^3\,\lsol)^{-0.3}$\ found by \citet{cabrit1992}, and later verified and extended by \citet{wu2004} for a larger sample of molecular outflows from both low and high-mass YSOs. This suggests that small-scale molecular disk winds are, at least energetically, indistinguishable from larger-scale molecular outflows, which are usually interpreted as swept-up material \citep{pascucci2023}.


\subsection{Comparison to other rotation detections of jets and disk winds} \label{ssec:dis:comp}

\begin{table*}[ht]
\caption{Comparison of rotating molecular winds\tablefootmark{a}.}
\label{tab-ofcomp} 
\begin{tabular}{llllllll}
\hline \hline
Source & Dist. & Age & $M_{\ast}$ & $R_{\rm L}$ & $j$ & $\lambda$\tablefootmark{b} & Reference \\
 & [pc] & [Myr] & [\msol] & [au] & [au\,\kms] & -- & \\
\hline 
HH\,211          & ~280 &  0.02         & 0.05         & 0.014      & 50        & $\ldots$ & \citet{lee2009,lee2018} \\
HH\,212          & ~400 &  0.04         & 0.25         & 0.05       & 10        & 2\,--\,6 & \citet{lee2017,tabone2017} \\
IRAS\,21078+5211 & 1630 &  $\ldots$     & $\approx$5.6 & $\le$6\,--\,17  & $\ldots$  & 2.4 & \citet{moscadelli2022} \\
HD\,163296       & ~101 &  $\approx$5   & 2.2          & $\ldots$   & $\ldots$  & $\ldots$ & \citet{klaassen2013} \\
DG\,Tau\,b       & ~140 & $\approx$1   &  1.1    & 2\,--\,5     & $\approx$65 & 1.6 & \citet{zapata2015,devalon2020} \\
SVS13A           & ~235 &  $<1$         & 0.9          & 7.5        & 6000      & $\ldots$ & \citet{chen2016} \\
NGC1333\,IRAS4C  & ~235 &  $<1$         & $\approx$0.2 & 5\,--\,15     & $120-210$ & 2\,--\,6 & \citet{zhang2018} \\
Orion\,I         & ~418 &  $\ldots$     & $\approx$5   & $>$10      & 500       & $\ldots$ & \citet{hirota2017} \\
TMC\,1A          & ~140 &  0.5          & 0.5          & 25         & 200       & $\ldots$ & \citet{bjerkeli2016} \\
CB\,26-YSO\,1    & ~140 &  $\approx$1 & 0.55         & 20\,--\,45         & 140       & 1\,--\,2 & this paper \\
\hline
\end{tabular} \tablefoot{
\tablefoottext{a}{Ordered by launch radius $R_{\rm L}$.}}
\tablefoottext{b}{Magnetic lever arm parameter $\lambda=\left(\frac{R_{\rm L}}{R_{\rm A}}\right)^2$, where $R_{\rm A}$\ is the Alfv{\`e}n radius \citep{blandford1982,ferreira2006}.}
\end{table*}

Rotation of jets and outflows has long been predicted theoretically, but observational evidence started accumulating only over the past 20 years. Rotation signatures in optical jets have been observed in a few sources and could be interpreted as tracing the helicoidal structure of the magnetic field in accretion-ejection structures \citep[e.g.,][]{bacciotti03,coffey04,coffey2007,woitas_etal2005}. 

The first reports of rotation signatures in slow molecular winds \citep{pascucci2023} started to emerge about 15 years ago. 
\citet{lee2009} report the discovery of rotation signatures in the low-velocity molecular outflow around the highly collimated jet from the $\sim$50\,\mjup\ BD-mass, $\sim2\times10^4$\,yr old protostar HH\,211 located in the IC\,348 complex of Perseus. Using Anderson's relation \citep{Anderson03}, they deduce an upper limit of the launching radius of $\sim$0.014\,au, which is much smaller than what we find in CB\,26, and conclude that the wind is consistent with being launched by magneto-centrifugal force. They estimate that the flow has a mean specific angular momentum of $\sim$50\,au\,\kms.
At about the same time, we reported in \citet{launhardt09} the discovery of rotation in the molecular outflow from the low-mass ($\sim$0.5\,\msol), $\sim$1\,Myr old Class\,I YSO CB\,26\,-\,YSO\,1.

\citet{klaassen2013} report the discovery of a rotating molecular disk wind from the young (4-7\,Myr) Herbig\,Ae star HD\,163296. They derive a mass loss rate that is twice as high as the accretion rate onto the star, but no constraints on the launch radius. 
\citet{chen2016} report the discovery of rotating molecular bullets associated with the high-velocity jet from the young variable Class 0/I transition protostar SVS\,13\,A located in the NGC\,1333 star-forming region. Using Anderson's relation \citep{Anderson03}, they deduce that the jet-launching footprint on the disk has a radius of $\sim$7.5\,au.
\citet{bjerkeli2016} report the discovery of a rotating molecular disk wind that is ejected from a region extending up to a radial distance of 25\,au around the young (few $10^5$\,yrs) low-mass ($\sim$0.5\,\msol) protostar TMC1A located in the Taurus molecular cloud. They derive a specific angular momentum of the outflowing gas of \mbox{$j\approx200$\,au\,\kms} that is slowly increasing with distance from star. This source is thus the one that is most similar, both in terms of launch radius and specific angular momentum, to the CB\,26 disk wind we report in this paper.
\citet{hirota2017} report signatures of rotation in the bipolar outflow driven by Orion Source\,I, a high-mass (5-6\,\msol) YSO candidate. They derive launching radii of $>$10\,au and an outward outflow velocity of $\sim$10\,\kms\ and conclude that the rotating outflow must be directly driven by a magneto-centrifugal disk wind.
\citet{zhang2018} report the discovery of clear signatures of rotation from about 120\,au up to about 1400\,au above the disk midplane of the low-mass ($\sim$0.2\,\msol) protostar NGC\,1333\,IRAS\,4C in the Perseus Molecular Cloud. From the angular momentum distribution in the outflow, they deduce the most likely launching radii to be 5–15\,au.

A break-through discovery was probably the detection of rotation in the high-velocity protostellar jet launched at \mbox{$R_{\rm L}=0.05$\,au} from the edge-on 0.04\,Myr old, $\sim$0.25\,\msol\ Class\,0 protostar HH\,212 in the L\,1630 cloud of Orion by \citet{lee2017}. 
In a series of succeeding papers, \citet{lee2021} analyze the interaction between the outer disk wind and the jet, and \citet{lee2022} find that the jet is the densest part of a wide-angle wind that flows radially outward at distances far from the (small, sub-au) launching region around the protostar HH\,212.

Another key observation was recently published by \citet{moscadelli2022}, who used observations at 22\,GHz with the VLBI network to obtain very high-spatial-resolution astrometry of water masers that are associated with the jet originating from a massive ($\approx$5.6\,\msol) YSO embedded in the IRAS\,21078+5211 star-forming region at a distance of 1.63$\pm$0.05\,kpc. They could show that the masers trace the velocities of individual streamlines spiraling outward along a helical magnetic field, launched from locations on the disk at radii $\le6-17$\,au, and that their motion is consistent with a MHD disk wind.

Table\,\ref{tab-ofcomp} summarizes the main parameters of the rotating molecular winds described here, approximately ordered by their respective launch radius $R_{\rm L}$. 
For comparison with CB\,26, we also list the respective values for the specific angular momentum of the outflown gas, $j$, and the magnetic lever arm parameter, $\lambda$, where these were explicitly given.
While the rotating jets associated with the youngest (few $10^4$\,yrs) low-mass Class\,0 protostars HH\,211 and HH\,212 are launched from within a few stellar radii with specific angular momenta of a few tens au\,\kms, the outflows from the somewhat older ($\sim$1\,Myr) Class\,I sources are driven by more extended molecular disk winds launched at radii of a few tens of au and specific angular momenta of a few hundred au\,\kms. The disk wind from \mbox{CB\,26-YSO\,1} appears to be the one with the largest launch radius, consequently also the best-resolved one, albeit quite comparable to, for example, TMC\,1A.


\section{Summary and conclusions} \label{sec:sum}

We obtained with the IRAM PdBI and analyze high angular-resolution (0\farcs53$\times$0\farcs47) $^{12}$CO(2--1) molecular line and thermal dust continuum data at 230\,GHz with the goal of revealing the nature and origin of the rotating molecular outflow from the young ($1_{-0.5}^{+2}$\,Myr) edge-on-seen \mbox{($i =87\pm2$\degr)} low-mass (0.55$\pm$0.1\,\msol) YSO\,\--\,protoplanetary disk system \mbox{CB\,26\,-\,YSO\,1}. The flared disk has an outer radius of \mbox{$R_{\rm out}^{\rm dust}\approx200$\,au}, an inner hole in dust emission of \mbox{$R_{\rm in}^{\rm dust}\approx16^{+37}_{-8}$\,au}, and a total mass of \mbox{$M_{\rm disk}\approx0.15\pm0.05$\,\msol}. The size of the inner hole in the CO emission of \mbox{$R_{\rm in}^{\rm CO}\approx12\pm6$\,au}, which we derive from a least-squares fit of our disk CO model to the channel maps,  might actually be an over-estimation and the CO disk might have a much smaller or even no inner hole at all. The source is embedded in the Bok globule CB\,26 \citep{launhardt09}. This relative isolation certainly favors both the undisturbed morphology and the good observability of the described disk wind. Furthermore, only the fact that the source is still embedded in its birth cloud enables the possibility of ongoing accretion onto the disk, which in turn could trigger such a powerful disk wind. Since the source is located too far north to be accessible to ALMA, the IRAM PdBI provided the best opportunity to obtain such high-resolution millimeter data.

Our observations confirm the disk-wind nature suggested by \citet{launhardt09} and reject the alternative scenarios such as jet precession or two misaligned jets from a hypothetical embedded binary system that were mentioned in the same paper. The new high-resolution data reveal an X-shaped morphology of the CO emission close to the disk, and vertical streaks extending from the disk surface out to vertical heights of $\approx$600\,au below, and $\approx$300\,au above. We interpret this emission as the combination of the disk atmosphere (the X-shaped part close to the disk) and a well-collimated disk wind, of which we mainly see the outer walls of the outflow cone. We decompose these two contributions by subtracting a chemo-dynamical model of the CO emission from the disk atmosphere combined with line radiative transfer calculations, fit to the data, from the channel maps of the observed CO emission. This allows us to trace the disk wind down to vertical heights of $\approx$40\,au where it is launched from the surface of the flared disk at a mean radius of $R_{\rm L}=$20\,--\,45\,au. The full launch region may actually extend over the entire inner disk from the inner rim at 10-30\,au out to the CO iceline at $\approx$80\,au.

The disk wind is rotating with the same orientation and speed as the Keplerian disk and the velocity structure of the cone walls, which open slowly with height (from \mbox{$R\approx$45\,au} at $|z|=40$\,au to $R\approx$100\,au at $|z|=500$\,au, with a half-opening angle $\theta=7\pm2\degr$) is consistent with angular momentum conservation of a flow that is launched from the disk surface at the single radius $R_{\rm L}=35\pm10$\,au with the Keplerian rotation velocity of \mbox{$V_{\rm K}=3.7\pm0.3$\,\kms}. From the \mbox{$^{12}$CO(2--1)} total intensity maps, we derive a total half-length of the CO-outflow of $r_0\approx1100$\,au and a total gas mass of \mbox{$M_{\rm CO}\approx(1\pm0.3)\times10^{-3}$\,\msol}. 
With a radial lift-off velocity of \mbox{$V_0=7_{-5}^{+20}$\,\kms}, we derive a dynamical age of the observed CO outflow of \mbox{$\tau_{\rm dyn}=740_{-500}^{+2000}$\,yr}, which does most likely not reflect the actual age of the outflow. A Herbig-Haro object (HH\,494) with a tangential velocity of 86$\pm$17\,\kms, located at 6\dotmin15 toward the north west, and exactly aligned with extended axis of the CO outflow at P.A.=148$\pm$1\degr, is suggested to originate from an accretion event that created an energetic inner disk wind and launched a yet undetected high-velocity jet about 3000\,yrs ago.

The currently observed outer disk wind is found to very efficiently ($\lambda\approx 1-2$) carry away angular momentum at a rate of $\dot{J}_{\rm CO}\approx 3\times10^4$\,\msol\,km$^2$\,s$^{-1}$\,yr$^{-1}$, which implies that the angular momentum flux by accretion onto the disk, for which we have no direct observational constraints, must be of the same order.
The wind has a total outflow momentum flux (thrust) of \mbox{$\dot{P}_{\rm CO}=\left(1.0_{-0.9}^{+2.6}\right)\times10^{-5}$\,\msol\,\kms\,yr$^{-1}$}, which is nearly three orders of magnitude larger than the maximum thrust that can be provided by a central star with bolometric luminosity \mbox{$L_{\ast}=(1\pm0.4)$\,\lsol}. Therefore, photoevaporation cannot be the main driving mechanism for this outflow, but it must be predominantly a MHD-driven disk wind. Indeed, the CB\,26 outflow falls right on the relation \mbox{$F_{\rm CO}/F_{\rm rad}=250\,(L_{\rm bol}/10^3\lsol)^{-0.3}$} found by \citet{cabrit1992} for a large number of molecular outflows over a wide range of luminosities.
It is thus far the best-resolved rotating disk wind observed to be launched from a circumstellar disk in Keplerian rotation around a low-mass YSO, albeit also the one with the largest launch radius. It confirms the observed trend that disk winds from Class\,I YSO's with transitional disks have much larger launch radii, and also larger specific angular momenta, than the inner disk winds and jets ejected from Class\,0 protostars.


\begin{acknowledgements}
We wish to thank Ralph Pudritz for inspiring discussions and very helpful comments on the manuscript. We also thank Christian Fendt and Ilaria Pascucci for helpful comments and discussions, and Uma Gorti for her contributions in the early stages of this project. Special thanks to Thomas M\"uller for his help in creating the cartoon of Fig.\,\ref{fig_sketch}.
We acknowledge the Plateau de Bure IRAM staff for their help during the observations. We also acknowledge the Grenoble IRAM staff for their support during the first data reduction session in January 2006. The French Program of Physico-Chemistry (PCMI) is thanked for providing fundings to this project. 
T.H. acknowledges support from the European Research Council under the Horizon 2020 Framework Program via the ERC Advanced Grant Origins\,83\,24,28.
Y.P. and V.A. were supported by the RSF grant 22-72-10029.
This publication makes use of data products from the Wide-field Infrared Survey Explorer (WISE), which is a joint project of the University of California, Los Angeles, and the Jet Propulsion Laboratory/California Institute of Technology, funded by the National Aeronautics and Space Administration. IRAM is supported by INSU/CNRS (France), MPG (Germany) and IGN (Spain). The Owens Valley millimeter-wave array was supported by NSF grant AST 9981546. The SMA is a joint project between the Smithsonian Astrophysical Observatory and the Academia Sinica Institute of Astronomy and Astrophysics and is funded by the Smithsonian Institution and the Academia Sinica. 
\end{acknowledgements}


\bibliography{cid_cb26-bib}
\bibliographystyle{aa}



\begin{appendix} 

\section{Channel maps}          \label{sec-app-chmaps}

Here we show the channel maps of $^{12}$CO\,(2--1) as observed, as modeled for the disk atmosphere, and observed minus disk model, as well as of $^{13}$CO\,(1--0), HCO$^+$\,(1--0), and HCO$^+$\,(3--2). All channel maps are overlaid with contours of the 1.3\,mm dust continuum emission from the disk.


\begin{figure*}
\begin{center}
\includegraphics[width=0.9\textwidth]{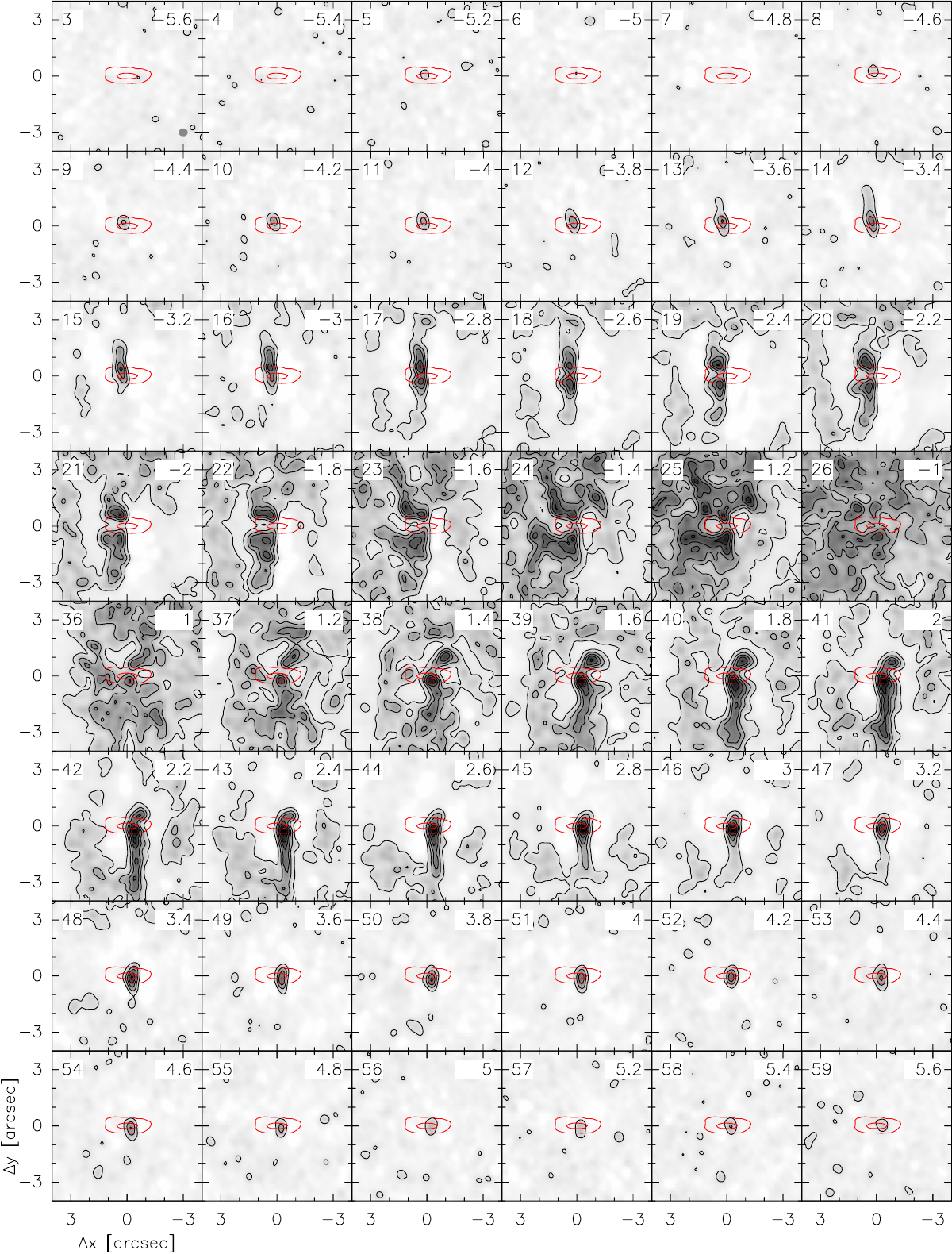}
\caption{\label{fig_chanmap_12co_obs}
$^{12}$CO\,(2--1) channel maps of CB\,26, obtained with PdBI in 2005 and 2009, rotated counterclockwise by $32\degr$.
Contour levels start at 15\,mJy/beam (2\,$\sigma$\ r.m.s., see Table\,\ref{tab-obs}). 
Red contours mark the 1.2\,mm dust continuum emission from the disk (3 and 15\,mJy/beam. 
The reference position is $\alpha_{2000} = 04^h59^m50.74^s$, 
                          $\delta_{2000} = 52^{\circ}04^{\prime}43.80^{\prime\prime}$.
The $^{12}$CO synthesized beam size is indicated as the gray ellipse in the lower right
corner of the first channel map. The channel number and mean $\Delta v = v_{\rm LSR}-5.95$\ 
(in \kms) are indicated in the top left and right corners of the maps, respectively. 
Maps in the central velocity channels between $\Delta v\approx\pm1.2$\,\kms are corrupted by 
resolved-out emission and self-absorption from the extended envelope.
}
\end{center}
\end{figure*}

\begin{figure*}
\begin{center}
\includegraphics[width=0.9\textwidth]{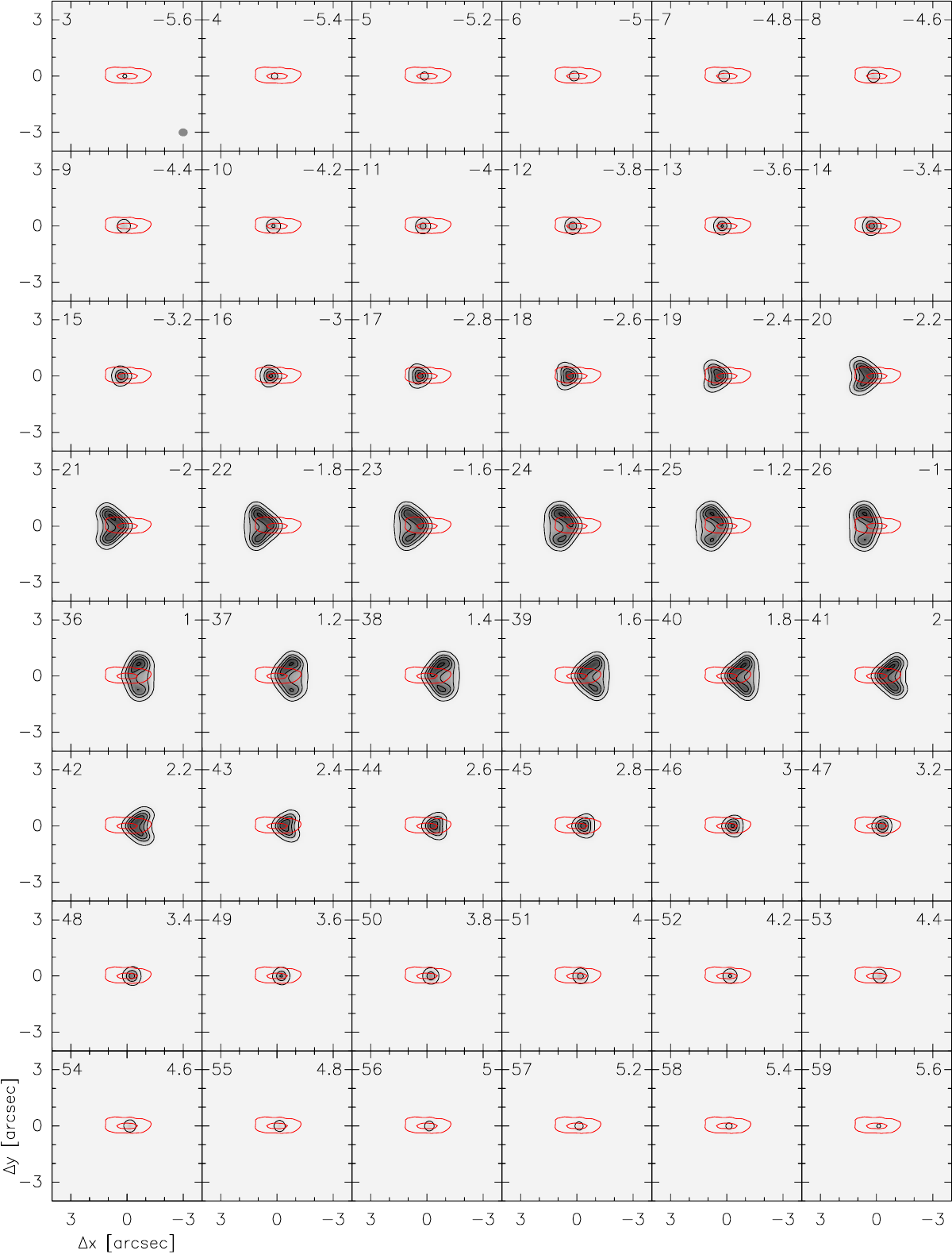}
\caption{\label{fig_chanmap_12co_mod}
Same as Fig.\,\ref{fig_chanmap_12co_obs}, but for the modeled $^{12}$CO\,(2--1) emission 
of the best-fit disk model for CB\,26 (see Table\,\ref{tab-diskpar}).
Contour levels start at 10\,mJy/beam. 
}
\end{center}
\end{figure*}

\begin{figure*}
\begin{center}
\includegraphics[width=0.9\textwidth]{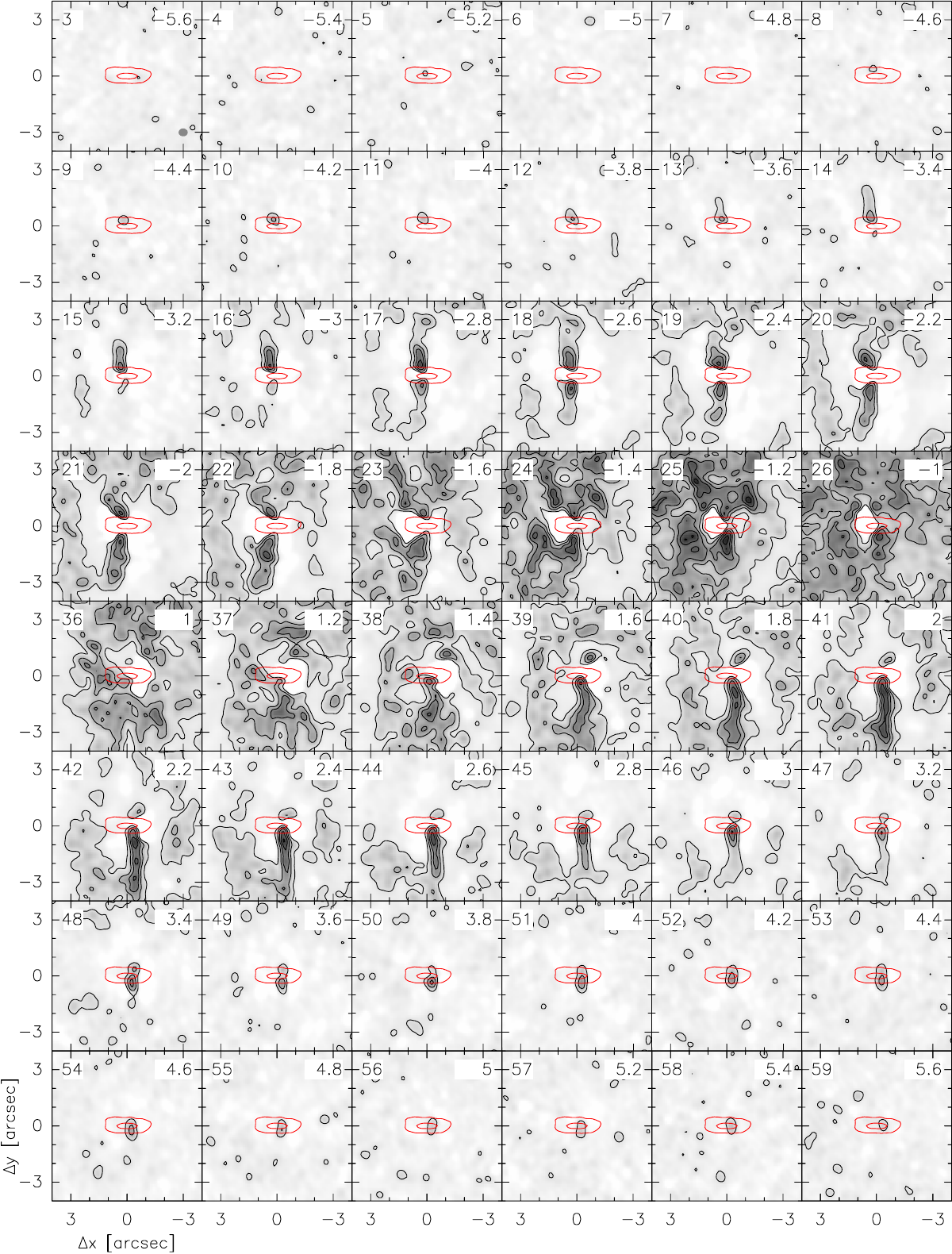}
\caption{\label{fig_chanmap_12co_diff}
Same as Fig.\,\ref{fig_chanmap_12co_obs}, but for the 
residual $^{12}$CO\,(2--1) emission from CB\,26 after subtracting the best-fit disk model 
(Fig.\,\ref{fig_chanmap_12co_mod}). Contour levels start at 15\,mJy/beam. 
}
\end{center}
\end{figure*}

\begin{figure*}
\begin{center}
\includegraphics[width=0.9\textwidth]{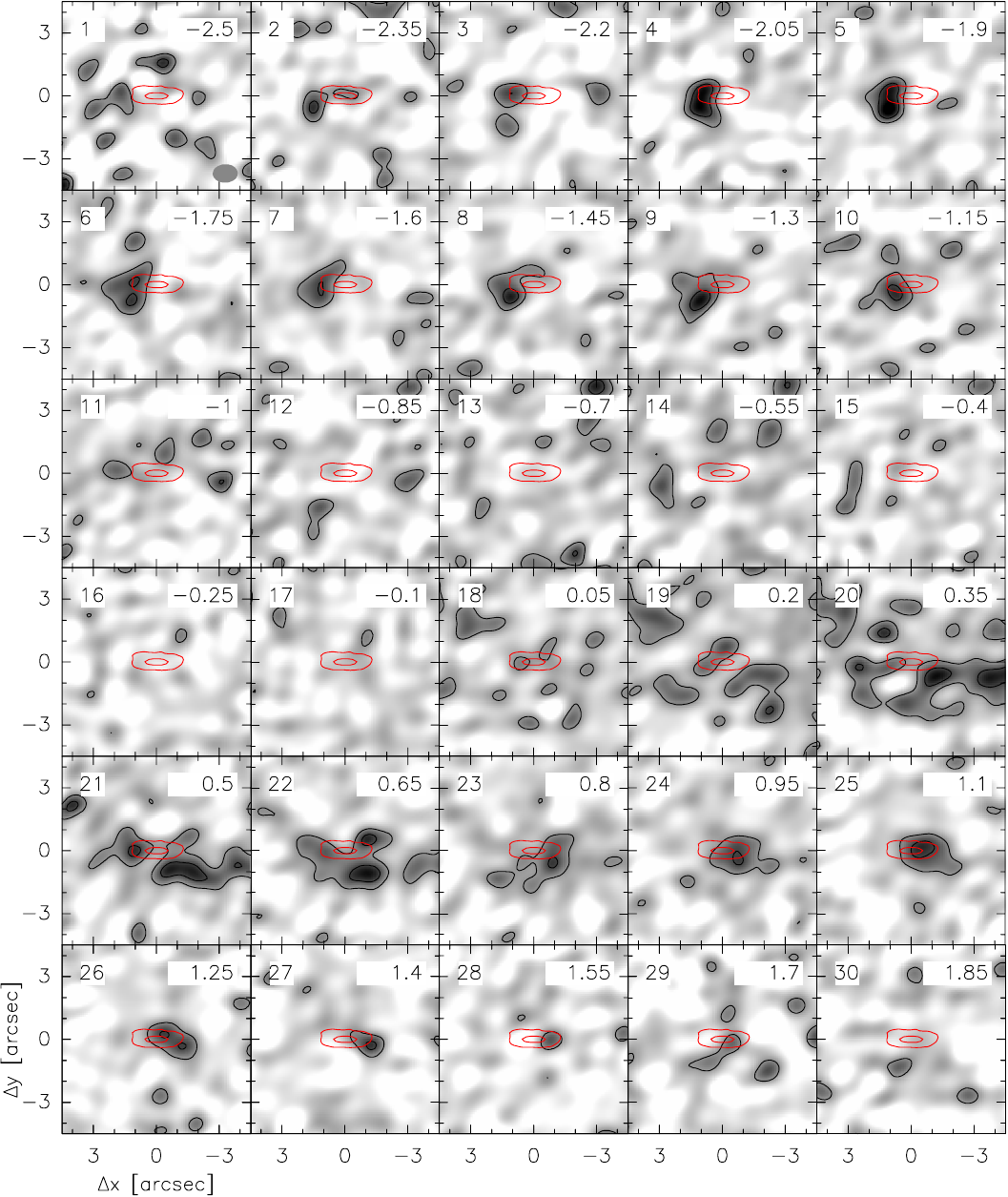}
\caption{\label{fig_chanmap_13co}
Same as Fig.\,\ref{fig_chanmap_12co_obs}, but for $^{13}$CO\,(1--0), obtained with OVRO in 2001. 
Contour levels start at 70\,mJy/beam (2\,$\sigma$\ r.m.s., see Table\,\ref{tab-obs}).
Maps in the central velocity channels between $\sim\pm1.2$\,\kms\ are corrupted by 
resolved-out emission and self-absorption from the extended envelope.
Maps in the central velocity channels between $\Delta v\approx -1.0$\ and 0.5\,\kms are corrupted by 
resolved-out emission and self-absorption from the extended envelope. 
}
\end{center}
\end{figure*}

\begin{figure*}
\begin{center}
\includegraphics[width=0.9\textwidth]{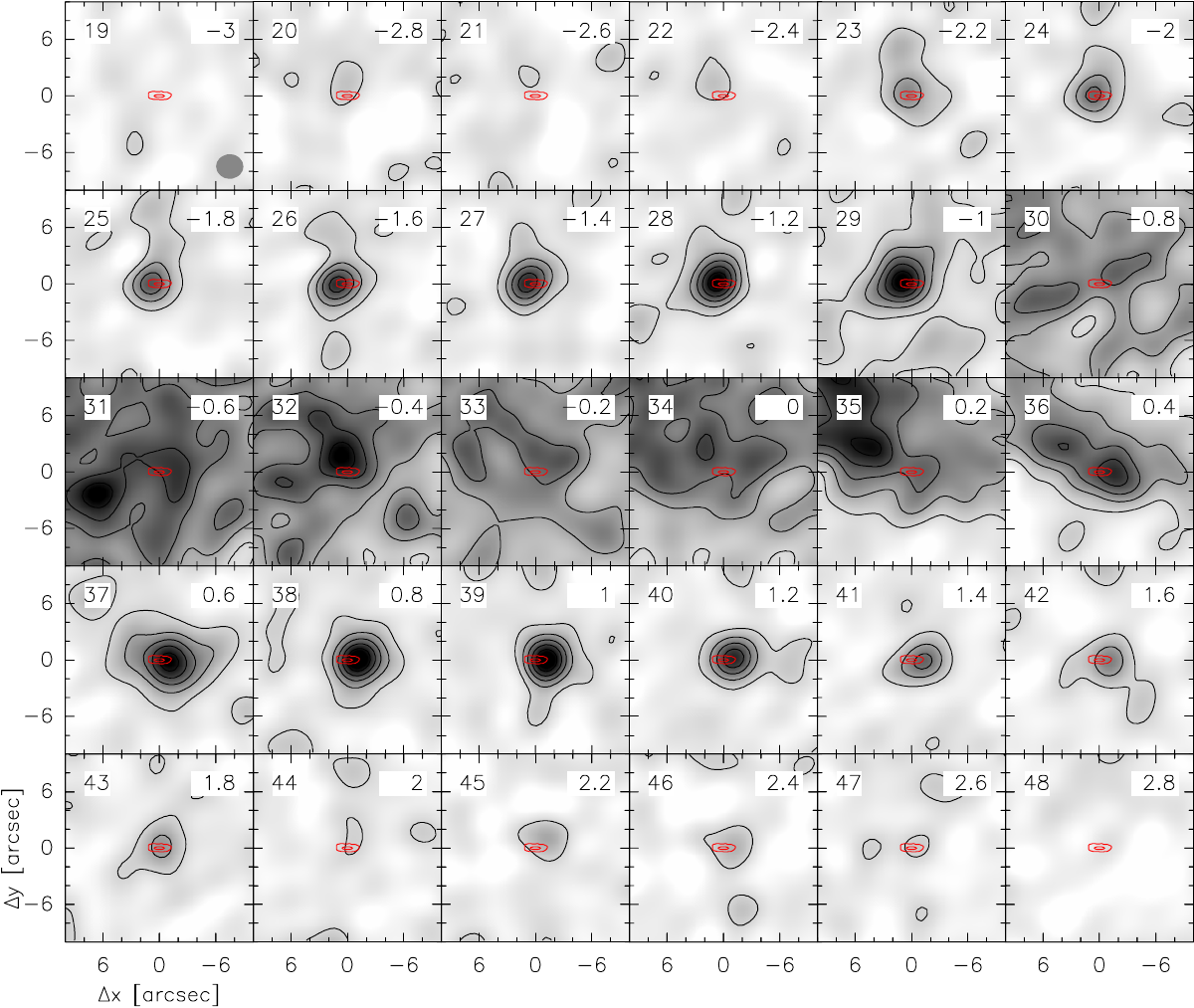}
\caption{\label{fig_chanmap_hco10}
Same as Fig.\,\ref{fig_chanmap_12co_obs}, but for HCO$^+$\,(1--0), obtained with PdBI in 2005. Contour levels start at 15\,mJy/beam ($\sim2\,\sigma$\ r.m.s., see Table\,\ref{tab-obs}). Maps in the central velocity channels between $\Delta v\approx -0.8$\ and 0.4\,\kms are corrupted by resolved-out emission and self-absorption from the extended envelope. 
}
\end{center}
\end{figure*}

\begin{figure*}
\begin{center}
\includegraphics[width=0.9\textwidth]{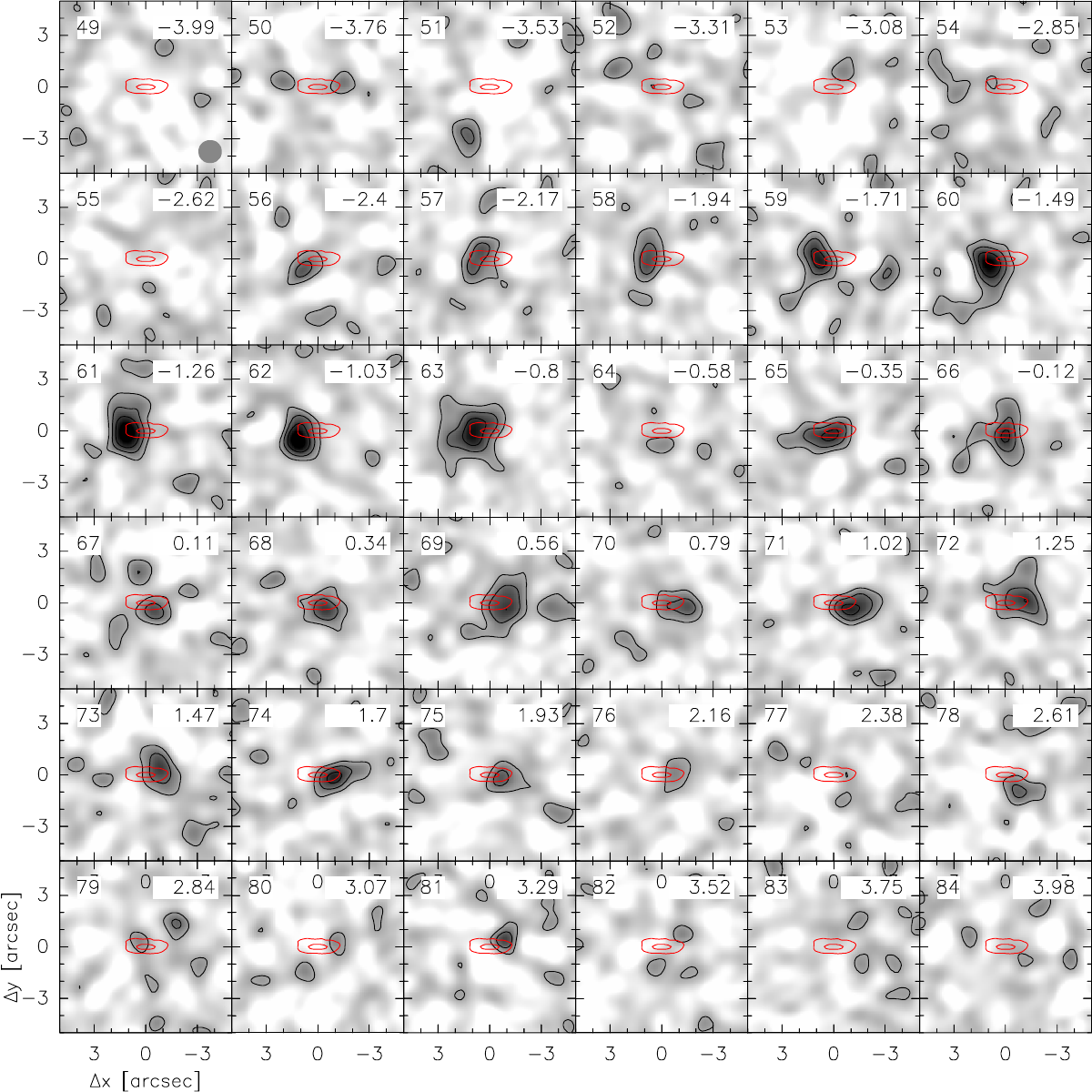}
\caption{\label{fig_chanmap_hco32}
Same as Fig.\,\ref{fig_chanmap_12co_obs}, but for HCO$^+$\,(3--2), obtained with the SMA in 2006. Contour levels start at 300\,mJy/beam (2\,$\sigma$\ r.m.s., see Table\,\ref{tab-obs}).
}
\end{center}
\end{figure*}


\section{The luminosity of the central star} 
\label{sec:app:lstar}

Integrating the SED (Fig.\,\ref{fig:sed}) and assuming isotropic irradiation and a distance of 140\,pc, yields a total luminosity of $\approx$0.43\,\lsol. However, this value may not correctly reflect the intrinsic bolometric luminosity of the central YSO since, in particular at the shorter wavelengths, anisotropic irradiation due to the nonspherical geometry becomes significant. We therefore also derive separately the thermal contribution to the luminosity at $\lambda>12\,\mu$m, which is less affected by the nonspherical geometry, and where most of the observed luminosity originates. In this wavelength range, we derive a total thermal luminosity of \mbox{$L_\mathrm{therm}^\mathrm{tot}\approx 0.42\,\mathrm{L}_{\odot}$}, of which \mbox{$L_\mathrm{therm}^\mathrm{disk}\approx 0.24\,\mathrm{L}_{\odot}$}\ (58\%) can be attributed to the disk and \mbox{$L_\mathrm{therm}^\mathrm{env}\approx 0.18\,\mathrm{L}_{\odot}$}\ (42\%) to the envelope. The value of \mbox{$L_\mathrm{therm}^\mathrm{disk}\approx 0.24\,\mathrm{L}_{\odot}$} is probably a good approximation, albeit slight underestimation, of the fraction of stellar luminosity that is intercepted and reprocessed by the disk.

Due to the nonspherical geometry and extreme edge-on configuration, the shorter-wavelength contribution to the luminosity from scattered light at $\lambda\le12\,\mu$m is likely to be significantly underestimated \citep[see, e.g.,][]{whitney2003}. Comparison with radiative transfer models from the grid of SED models of YSOs provided by \citet{robitaille2006} for configurations similar to CB\,26 (e.g., model-ID 3016525) suggests that the NIR to MIR luminosity could be underestimated by of up to a factor of ten and the total luminosity by up to a factor of two. Vice versa, for a similar system seen pole-on, the observed luminosity could overestimate $L_{\rm bol}$\ by up to a factor of two. We therefore estimate the most likely value of the intrinsic bolometric luminosity of the central star(s) to be $\tilde{L}_{\rm bol} = 1.0 \pm 0.4$\,\lsol. This value is in good agreement with \citet{launhardt09} and \citet{zhang2021}, who both estimated $L_{\ast}\approx0.9$\,\lsol.


\section{Proper motion of HH\,494} \label{sec:app:hh494}

The precise alignment of HH\,494 at \mbox{P.A.\,$=147.5\pm0.5$\degr}\ with regard to CB\,26-YSO\,1 \citep{stecklum04} 
with the axes of the CB\,26 disk and outflow (\mbox{P.A.\,$=148\pm1$\degr}, Sect.\,\ref{ssec:res:ov}, see also Fig.\,\ref{fig_intmaps1}) suggests that this HH object may originate from a jet that was ejected from CB\,26\,-\,YSO\,1. Although no such jet nor another HH object have been detected yet, part of the explanation for the location of HH\,494 may come from the specific environment of CB\,26. As indicated by the diffuse cloudshine\footnote{Stellar photons that are scattered at small dust grains in diffuse clouds or in the optically thin halos around dense cores and filaments \citep[e.g.,][]{foster2006}.} (Fig.\,\ref{fig_intmaps1}), the vicinity of the Bok globule is relatively void of extincting and light-scattering material. HH\,494 is located where the connecting line with \mbox{CB\,26\,-\,YSO\,1} encounters the first layer of diffuse material. It may thus mark the interaction spot of the (currently unseen) jet with the first layer of material it encountered after leaving the boundaries of the globule.

Based on the data described in Sect.\,\ref{sec:obs:nir}, we also derived the proper motion vector of HH\,494. Since the position of the HH object is identical on the H$\alpha$ and S[{\sc ii}] images of both epochs within the uncertainties, we co-added them to increase the signal-to-noise ratio (S/R). Because the HH object is located close to a star of comparable brightness, which might affect the proper motion analysis, the continuum emission was removed by subtracting the properly scaled $R$-band frame, after accounting for a fractional pixel shift. On the resulting images, object detection was performed using the SEXTRACTOR code \citep{bertin1996} at the 3\,$\sigma$\ detection level. Thereby, the positions of HH\,494 for the two epochs were obtained with coordinate uncertainties of 0\farcs12\,--\,0\farcs17. From these measurements, coordinate differences of \mbox{$\Delta$RA=0\farcs89$\pm$0\farcs17} and \mbox{$\Delta$DEC=1\farcs29$\pm$0\farcs22} were derived. At a distance of 140\,pc and for the epoch difference of 12.1478\,yrs, these translate into a tangential velocity of \mbox{86$\pm$17\,\kms}. Since the inclination of the disk and outflow axes is very close to 90\degr\ (see Sect.\,\ref{ssec:mod:disk} and Table\,\ref{tab-diskpar}), this tangential velocity should be very close (within 2\%) to the actual space velocity. The position angle of the proper motion of 325\fdg3$\pm$4\fdg8 (E of N) is consistent (within the uncertainties) with the orientation of the bipolar nebula and the molecular outflow ($148+180\Rightarrow 328\pm 1\degr$, Sect.\,\ref{ssec:res:ov}). This excellent agreement supports our hypothesis that HH\,494 originates from a jet that was ejected from CB\,26\,-\,YSO\,1. 

With a distance of 140$\pm$20\,pc and a very close to edge-on orientation, the projected separation of HH\,494 from CB\,26-YSO\,1 of 6\dotmin15\ translates into a physical separation of $\approx 5\times10^4$\,au or 0.25\,pc If we now assume that {\it(i)} this jet travelled at constant speed and {\it(ii)} that HH\,494 has preserved the jet velocity, the corresponding dynamical time scale would $2860\pm360$\,yrs. Since neither of these two assumptions, in particular the second one, may be correct, this dynamical time scale is only a very crude estimate of the age of the corresponding ejection event and the actual uncertainty may be much larger than our formal error. We discuss the context and implications of this estimate in Sect.\,\ref{ssec:dis:outflow}.

\end{appendix}


\end{document}